\documentstyle[epsfig,longtable,times]{mn}

\newif\ifAMStwofonts

\ifoldfss

  \renewcommand{\chem}[2] {$\rm{}^{#2}\kern-0.8pt#1$}
  \ifCUPmtlplainloaded \else
    \NewTextAlphabet{textbfit} {cmbxti10} {}
    \NewTextAlphabet{textbfss} {cmssbx10} {}
    \NewMathAlphabet{mathbfit} {cmbxti10} {} 
    \NewMathAlphabet{mathbfss} {cmssbx10} {} 
  \fi
  \ifAMStwofonts
    \ifCUPmtlplainloaded \else
      \NewSymbolFont{upmath} {eurm10}
      \NewSymbolFont{AMSa} {msam10}
      \NewMathSymbol{\upi}     {0}{upmath}{19}
      \NewMathSymbol{\umu}     {0}{upmath}{16}
      \NewMathSymbol{\upartial}{0}{upmath}{40}
      \NewMathSymbol{\leqslant}{3}{AMSa}{36}
      \NewMathSymbol{\geqslant}{3}{AMSa}{3E}

       \let\le=\leqslant
      \let\geq=\geqslant \let\ge=\geqslant
    \fi
  \fi
\fi 

\ifnfssone
  \newmathalphabet{\mathit}
  \addtoversion{normal}{\mathit}{cmr}{m}{it}
  \addtoversion{bold}{\mathit}{cmr}{bx}{it}
  \newmathalphabet{\mathbfit} 
  \addtoversion{normal}{\mathbfit}{cmr}{bx}{it}
  \addtoversion{bold}{\mathbfit}{cmr}{bx}{it}
  \newmathalphabet{\mathbfss} 
  \addtoversion{normal}{\mathbfss}{cmss}{bx}{n}
  \addtoversion{bold}{\mathbfss}{cmss}{bx}{n}
  \ifAMStwofonts
    \ifCUPmtlplainloaded \else
      \UseAMStwoboldmath
      \makeatletter
      \new@mathgroup\upmath@group
      \define@mathgroup\mv@normal\upmath@group{eur}{m}{n}
      \define@mathgroup\mv@bold\upmath@group{eur}{b}{n}
      \edef\UPM{\hexnumber\upmath@group}
      \new@mathgroup\amsa@group
      \define@mathgroup\mv@normal\amsa@group{msa}{m}{n}
      \define@mathgroup\mv@bold\amsa@group{msa}{m}{n}
      \edef\AMSa{\hexnumber\amsa@group}
      \makeatother
      \mathchardef\upi="0\UPM19
      \mathchardef\umu="0\UPM16
      \mathchardef\upartial="0\UPM40
      \mathchardef\leqslant="3\AMSa36
      \mathchardef\geqslant="3\AMSa3E

       \let\le=\leqslant
      \let\geq=\geqslant \let\ge=\geqslant
    \fi
  \fi
\fi 

\ifnfsstwo
  \DeclareMathAlphabet{\mathbfit}{OT1}{cmr}{bx}{it}
  \SetMathAlphabet\mathbfit{bold}{OT1}{cmr}{bx}{it}
  \DeclareMathAlphabet{\mathbfss}{OT1}{cmss}{bx}{n}
  \SetMathAlphabet\mathbfss{bold}{OT1}{cmss}{bx}{n}
  \ifAMStwofonts
    \ifCUPmtlplainloaded \else
      \DeclareSymbolFont{UPM}{U}{eur}{m}{n}
      \SetSymbolFont{UPM}{bold}{U}{eur}{b}{n}
      \DeclareSymbolFont{AMSa}{U}{msa}{m}{n}
      \DeclareMathSymbol{\upi}{0}{UPM}{"19}
      \DeclareMathSymbol{\umu}{0}{UPM}{"16}
      \DeclareMathSymbol{\upartial}{0}{UPM}{"40}
      \DeclareMathSymbol{\leqslant}{3}{AMSa}{"36}
      \DeclareMathSymbol{\geqslant}{3}{AMSa}{"3E}

       \let\le=\leqslant
      \let\geq=\geqslant \let\ge=\geqslant
    \fi
  \fi
\fi 

\ifCUPmtlplainloaded \else
  \ifAMStwofonts \else 
    \def\upi{\pi}
    \def\umu{\mu}
    \def\upartial{\partial}
  \fi
\fi

\title{Carbon Star Populations in Systems with Different 
       Metallicities: Statistics in Local Group Galaxies.}
\author[Mouhcine \& Lan\c{c}on]
       {M. Mouhcine, A. Lan\c{c}on\\
       Observatoire Astronomique de Strasbourg (UMR 7550),
       11, rue de l'Universit\'e, 67000 Strasbourg, France.}       
\date{Accepted ?.
      Received ?;
      in original form ?}

\pubyear{2001}

\begin{document}

\maketitle

\label{firstpage}

\begin{abstract}

We present evolutionary population synthesis models for the study  
of the cool and luminous intermediate age stellar populations in
resolved galaxies with particular emphasis on carbon star populations.
We study the effects of the star formation history, the age and
the metallicity on the populations of intermediate mass stars.
In the case of instantaneous bursts, we confirm that lower metallicity 
results in higher contributions of carbon stars to the total star number, 
and in higher number ratios of carbon stars to late-type M stars.

Chemically consistent models are used to study the effect of the star 
formation history on the relations between carbon star population 
properties and global parameters of the parent galaxy (age, metallicity). 
Our models are able to account, for the first time, 
for those correlations, as observed in the galaxies of the Local Group. 
For stellar populations older than about 1 Gyr, the properties of carbon 
star populations are linked to the current metallicity in a way that
is quite independent of the star formation 
scenario. The number ratio of carbon stars to late-type M stars forms
a metallicity sequence along which stellar populations with 
very different star formation histories are found.

For the same populations, we find that both the mean bolometric luminosity 
of carbon stars and their normalized number to the luminosity of the parent 
galaxy are quite independent of metallicity over a large range in metallicity.
This is in good agreement with the observational constraints.

The observed statistics of carbon star populations can be interpreted by 
two principal effects: (i) the carbon star formation efficiency is higher 
in metal-poor systems, (ii) the typical star formation timescale along the 
Hubble sequence of galaxies is much longer than the typical timescale for
the production of carbon stars at any metallicity.

\end{abstract}

\begin{keywords}
stars: AGB - stars: carbon - stars: evolution - stars: mass loss - 
galaxy: dwarf, morphology
\end{keywords}

\section{Introduction}
\label{intro.sec}
The stellar content of galaxies has long been recognized as holding 
important clues to the understanding of the formation and the evolution 
of galaxies.
Better understanding of galaxy evolution in terms of stellar populations 
and chemical evolution requires knowledge of the Asymptotic Giant Branch 
(AGB) stars. From a purely pragmatic point of view, the evolved AGB stars 
are easy targets to observe in external galaxies even far beyond the Local 
Group, and to segregate from the bulk of the stellar population: they are 
red and luminous sources. Stars in the AGB phase make a significant 
contribution to the integrated light of a stellar population (Frogel et 
al. 1990). Mouhcine \& Lan\c{c}on (2002) estimate that the thermally 
pulsing AGB stars (TP-AGBs) are responsible for 30--60\% of
the integrated JHK luminosities of a system whose stars are 
0.2-2\,Gyr old, with a maximum contribution at $\sim$\,0.8-1\,Gyr. 
Those properties make AGB stars useful tools to get information on the 
star formation (SF) history, even through several magnitudes of absorption 
(Hodge 1989). 
Due to their large luminosities, carbon stars of the AGB are used as 
tracers of kinematics to probe morphological and kinematical structures 
(Aaronson \& Olszewski 1987, Hardy et al. 1989, Kunkel et al. 1997,  
Graff et al. 2000).

Stars in the AGB phase are classified as either oxygen rich (C/O\,$<$\,1
by number) or carbon rich (C/O\,$\ge$\,1). 
Spectra of oxygen rich stars are dominated by metal oxyde bands such 
as TiO, VO, and H$_{2}$O whereas carbon stars have bands of 
C$_{2}$ and CN (Barnbaum et al. 1996, Joyce 1998). Using those 
features, groups led by Richer (Richer et al. 1984, Richer et al. 1985, 
Pritchet et al. 1987, Hudon et al 1989), Aaronson and co-workers 
(Aaronson et al. 1982, Aaronson et al. 1984, Cook et al. 1986, 
Aaronson et al. 1985, Cook \& Aaronson 1989) developed a technique 
to identify AGB stars and to determine their nature in crowded fields. 
This technique involves imaging a field though four filters. Two 
narrow-band filters provide spectral information on the CN and TiO 
bands (Wing 1971), while broad-band colours provide information on 
the effective temperature of the stars, and can discriminate between 
early type and late-type stars. Extensive observational work, using 
this technique and others (Frogel \& Richer 1983, Azzopardi et al. 1985, 
1986, 1998; Azzopardi \& Lequeux 1992; Westerlund et al. 1987, 1991\,a 
\& b, 1995; Demers et al. 1993, Battinelli \& Demers 2000) has lead to 
the identification and classification of AGB stars in nearby galaxies.
Recent comprehensive compilations of late-type stellar contents, and 
of their systematic statistics, can be found in Mateo (1998), 
Groenewegen (1999) and Azzopardi (2000). 

In order to interpret those findings, we have developed a chemically
consistent population synthesis model that includes the various
phenomena relevant to the production of carbon stars, and  their
dependence on metallicity. We use it to evaluate, qualitatively and 
quantitatively, (i) the interplay between different physical processes 
that operate during 
the AGB phase and control the formation of carbon stars, 
(ii) how these processes lead to the observed statistics of 
carbon star populations in the galaxies of the Local Group, and 
(iii) the sensitivity of the carbon star statistics on the global 
properties of the host galaxies (SF history, evolutionary status).
The paper is organized as follows. In Section 2 we present our current 
observational knowledge of the statistics of carbon star populations 
in the Local Group. In Section 3, we describe the calibrated 
semi-analytical evolution models that we use to follow the 
evolution of the TP-AGB stars. We also describe the 
population synthesis and the chemical evolution models. 
In Section 4 and 5, we present the predicted statistics of carbon 
stars for single stellar populations (SSPs) and address the effect 
of opaque dust envelopes on those statistics.
In Section 6 we calculate the statistics for continuous SF histories 
representative of various morphological types, 
and compare our results with the observational data.
In Section 7, we come back to the use of the relative number of 
carbon stars as an abundance indicator. The conclusions are
summarized in Section 8.

\section{Observational statistics and constraints}
\label{Obs.sec}

\subsection{The number ratio of carbon stars to late type M stars}
\label{NcNm.obs.sec}

Large sets of observational data on resolved  galaxies clearly show
that the number ratio of carbon stars to late type
M stars, N$_{C}$/N$_{M}$, depends on metallicity, in the sense that 
higher ratios are observed in lower metallicity environments 
(Blanco \& McCarthy 1983, Richer et al. 1985, Cook et al. 1986, 
Mould \& Aaronson 1985, Aaronson et al. 1987, Brewer et al. 1996, 
Albert et al. 2000). Even though large observational uncertainties 
affect metallicity and N$_{C}$/N$_{M}$ measurements (about 0.2 dex for 
[Fe/H] and 50\% for N$_{C}$/N$_{M}$ ratio), and although different 
carbon star surveys have covered different fractions 
of the area of the target galaxies with different completeness limits,
the N$_{C}$/N$_{M}$ vs. [Fe/H] correlation is well established 
(Groenewegen 1999). The observed sequence spans over a wide range in 
metallicity and N$_{C}$/N$_{M}$ (1.5 dex in [Fe/H] and 4 dex in 
N$_{C}$/N$_{M}$). The most striking feature of this 
sequence is that galaxies with different morphologies and different 
carbon star luminosity functions, indicating different SF histories, 
lie on the same sequence. This behavior suggests that the parent 
galaxy metallicity is the predominant factor, in comparison to age, 
in determining the observed ratio. 

Two arguments have been developed qualitatively in the literature 
to explain this effect. The first one is related to the effect of
metallicity on the effective temperature of the giant branch.
It leads to larger numbers of late type M stars in metal-rich systems.
The second one is a more efficient mixing of carbon from the
stellar core into the atmosphere in low-metallicity stars.
However, those qualitative arguments must be viewed with some
caution, since the SF history may also play some role in 
determining the relative frequency of carbon and M stars in 
galaxies.

\subsection{The mean bolometric luminosity of carbon rich stars}
\label{Lbolc.obs.sec}

The determination of the mean bolometric luminosity of the whole carbon 
star population is sensitive to distance determination, survey 
completeness limits, or surface coverage of the galaxy. Rapid examination 
of the data (Fig. 6 of Groenewegen 1999, or Fig. 9b of Aaronson 
\& Mould 1985) can lead one to find an anticorrelation between the 
mean bolometric magnitude of carbon star populations and the metallicity. 
However, some caution must be exerted.
Carbon stars in Ursa Minor or Draco for example, 
lie below the giant branch tip in the colour-magnitude 
diagram (Aaronson \& Mould 1985). They may not be on the AGB. 
Indeed there are two possible origins of these faint carbon stars
(see also Sect.\,\ref{Cstar_formation.sec}). 
The first one is that we are seeing dwarf 
carbon stars (i.e. not AGB stars), enriched in carbon by
mass transfer from a companion. 
The second possibility is that we are looking at carbon stars on the AGB, 
but evolving through the luminosity dip that follows a thermal pulse. 
The second possibility is unlikely  when {\em only} carbon stars
fainter than the tip of the red giant branch are found, as the luminosity
dips represent a small fraction of the thermal pulse cycle: carbon stars
with magnitudes consistent with the AGB core mass-luminosity relation
should also be present (see Sect.\,\ref{Models.sec}).

If one considers only those stellar populations whose carbon stars are 
in the AGB phase, the derived mean bolometric luminosity of the carbon 
star population seems
to be constant and equal to  $\approx\,-4.7$ over a large range of 
metallicities. This observational fact was reported in the literature 
since the beginning of the 80's. Aaronson \& Mould's data (1985) 
support this result. Richer et al. (1985) concluded that as long 
as a galaxy is relatively metal rich ([Fe/H]\,$>$\,-1.8), its carbon 
stars will have the same mean bolometric luminosity as those of Fornax, 
the Magellanic Clouds, the Milky Way, M31, and NGC 205. 
Brewer et al. (1996) observed different fields in M31, and they also 
derived that all have the same mean bolometric luminosity.  

One has to keep in mind that the universal quantity is the mean 
bolometric luminosity of the whole AGB carbon star population, 
not the carbon star luminosity function. The bright tail of the 
luminosity function is sensitive to recent star
formation episodes (Marigo et al. 1999). 

\subsection{The luminosity-normalized number of carbon stars}
\label{N_cL.obs.sec}

The third statistical property of carbon star populations 
is the behavior of the luminosity-normalized number of carbon stars 
as a function of galaxy metallicity. 
This quantity is a measurement of the number of carbon stars per V-band 
(or sometimes B-band) luminosity unit, grossly equivalent 
to a number of carbon stars per 
mass unit (Aaronson et al. 1983, Aaronson \& Mould 1985). 
We define $\log\,N_{\rm C,L}\,=\,\log\,N_{\rm Tot,C}$+0.4\,M$_{V}$,
where N$_{\rm Tot,C}$ is the total number of carbon stars and M$_V$ 
the integrated absolute V band magnitude of the whole population.   
N$_{\rm C,L}$ is of interest because it gives a possibility of 
deriving constraints on the physical processes that control the 
formation of carbon stars. 

The observational determinations of this quantity suffer from the
uncertainties in the surface photometry, internal absorption,
possible incompleteness, and errors
on distance determinations.
Most of the observational determinations of $\log\,N_{\rm C,L}$ 
scatter between a value of -3 and -4 up to [Fe/H]$\,\approx\,-0.5$. 
For higher metallicities, there is a large dispersion in N$_{\rm C,L}$ 
at constant metallicity, with evidence for a trend of decreasing 
N$_{\rm C,L}$ for increasing [Fe/H]. Azzopardi et al. (1999) noted 
that on average $\log\,N_{C,L}\,\simeq\,-3.3$ for Local Group galaxies 
over a large range in magnitude (or metallicity). Draco, Ursa 
Minor and Carina  stand out with large empirical values of N$_{\rm C,L}$ 
but, as mentioned in Sect.\,\ref{Lbolc.obs.sec}, the carbon stars in those 
galaxies may not be in the AGB phase. The observed behavior appears 
to exclude the interpretation of the N$_{C}$/N$_{M}$ vs. [Fe/H] 
relation as due only to the effect of the metallicity on the effective 
temperature of the red giant branch.
The relative insensitivity of N$_{C,L}$ to metallicity suggests that 
the {\em total} number of carbon stars scales with the parent galaxy 
magnitude, and consequently with the galaxy metallicity.
     
\section{Evolutionary Modeling}
\label{Models.sec}

\subsection{Stellar population models}
Our calculations are based on the stellar evolutionary tracks
of the Padova group (Bressan et al. 1993, Fagotto et al. 1994\,a,b).
For intermediate age stars, these tracks do not include TP-AGB phase.
We include that phase with new synthetic evolution models for 
metallicities [Z=0.0004, Y=0.23], [Z=0.004, Y=0.24], [Z=0.008, Y=0.25], 
[Z=0.02, Y=0.28], and [Z=0.05, Y=0.352].
In the following subsections, we highlight only the most important 
physical processes that affect the behavior of the TP-AGB stars in the
HR diagram and their chemical nature, as it was pointed out with detailed 
structural models (Boothroyd \& Sackmann 1992, Bl\"ocker \& Sch\"onberner 
1991, Frost et al. 1998, Wagenhuber \& Tuchman 1996).
Synthetic evolution models account for these effects with analytical 
formulae derived from full numerical calculations, and will remain an 
attractive tool as long as large enough grids of complete models for 
the structural evolution of TP-AGB stars stay out of reach (Iben \& 
Truran 1978, Renzini \& Voli 1981, Groenewegen \& de Jong 1993, Marigo 
et al. 1996). Complementary information on our models is given by 
Mouhcine \& Lan\c{c}on (2002), who focused on the integrated 
spectrophotometric evolution of intermediate age populations
at the metallicities of the solar neighborhood and the Magellanic
Clouds.

The assumptions on which TP-AGB evolution models stand are motivated 
mainly by the ability of their predictions to reproduce the observations
of simple stellar populations.
Mouhcine \& Lan\c{c}on (2002) and Mouhcine et al. (2002) show that the 
adopted ones adequately reproduce the emission properties of star clusters.
Here, our principal goal is to study resolved carbon star populations, 
and the sensitivity of their properties to star formation history and 
chemical evolution of galaxies. 
 
\subsection{TP-AGB evolution models}
\subsubsection{Dredge-up and Hot Bottom Burning}
The third dredge-up is of crucial importance for the formation of carbon 
stars. The adopted semi-analytical treatment of this process requires 
the knowledge of three inputs: (i) the critical core mass M$_{c}^{min}$ 
above which dredge-up is triggered, (ii) the dredge-up efficiency (defined 
as $\lambda=\Delta M_{dredge}/\Delta M_{c}$, where $\Delta M_{dredge}$ 
and $\Delta M_{c}$ are, respectively, the amount of mass dredged-up to 
the envelope after a pulse and the increase in core mass during the time 
between two successive TPs), and (iii) the element abundances in the 
inter-shell material. 
The theoretical values of the dredge-up parameters are uncertain. Standard 
numerical models predict low dredge-up efficiencies (Boothroyd \& Sackmann 
1988), while models assuming diffusive overshooting predict highly efficient
dredge-up events (Herwig et al. 1998). 
Actually both parameters $\lambda$ and M$_{c}^{min}$ affect the lifetime
of the carbon-rich phase and the age interval for which carbon stars are 
present in a stellar population. Both parameters are also expected to depend 
on properties such as the stellar metallicity or its instantaneous mass and 
structure, but the available information on these relations is neither
complete nor easy to extrapolate (e.g. Herwig 2000, Marigo et al. 1999).
Our models assume constant $\lambda$ and $M_c^{min}$. With this assumption, 
reproducing the luminosity function of carbon stars in the LMC requires 
intermediate dredge-up efficiencies and leads us to adopt $\lambda=0.75$ 
and $M_c=0.58$ (Groenewegen \& de Jong 1993, Marigo et al. 1996). The 
chemical composition of the dredged-up material is taken from Boothroyd 
\& Sackmann 1988. 

The hot bottom burning  (HBB, also known as envelope burning) is 
the second process affecting the chemical evolution of TP-AGB stars. 
This process is responsible for the destruction of newly dredged-up 
material at the base of the stellar envelope, in particular of 
carbon, and it thus prevents the formation of carbon rich stars 
in certain cases (Boothroyd et al. 1995). 
Envelope burning invalidates core mass-luminosity relations that 
were considered standard in the early 1990s (Bl\"ocker \& Sch\"onberner 
1991), and allows growth to higher luminosity (see Marigo et al. 1999b 
for a review). This triggers high mass loss rates, which in turn reduces 
the total lifetime of the TP-AGB phase and the final core mass.
The semi-analytical treatment of hot bottom burning requires to set 
the critical envelope mass necessary for ignition at the basis 
of the envelope. We have used the condition derived by Wagenhuber 
\& Groenewegen (1998). Combined with the mass loss prescription
described below, this ensures agreement with empirical initial-final
mass relations and with the age dependence of the TP-AGB contributions
to the light of star clusters (Mouhcine \& Lan\c{c}on 2002).

We have used the highly detailed analytical representations of 
Wagenhuber \& Groenewegen (1998) for both the core mass/luminosity, 
and core mass/interpulse period relations. Both third dredge-up and 
envelope burning were taken into account to calculate the evolution 
of the chemical abundance of stellar envelopes through the TP-AGB phase. 

\subsubsection{Mass loss}

Mass loss along the TP-AGB is represented by the prescription of
Bl\"{o}cker's (1995), which was derived from the hydrodynamical models 
of Long Period Variable stars of Bowen (1988):
\begin{equation}
\dot{M} = 4.83 \,\,10^{-9} \eta L^{2.7} M^{-2.1} \dot{M_{R}}
\end{equation}
$\eta$ is the mass loss efficiency, and $\dot{M_{R}}$ is Reimers's 
standard mass loss rate 
(i.e., $\dot{M_{R}}\,=\,\-1.27\,10^{-5}\, M^{-1}L^{1.5}T^{-2}_{eff}$). 
This formulation reflects the strong increase of the mass 
loss at the end of the AGB evolution, the so-called superwind phase.
The mass loss efficiency used is $\eta$\,=\,0.1, again as a result of 
calibration against carbon star populations of the LMC (Groenewegen 
\& de Jong 1994). 
$\eta$ is expected to depend on metallicity through variations
in the gas-to-dust ratio and in the resulting radiative acceleration,
but also through the essentially unknown metallicity dependence
of the LPV pulsation properties. In the absence of reliable relations,
we keep $\eta$ constant, and thus include no explicit metallicity 
dependence in the mass loss rate. The metallicity dependence appears 
only via the metallicity dependent location of giant stars in the HR 
diagram. 
Note that different mass loss prescription with appropriate mass 
loss efficiencies, within the framework of a self-consistent model 
of stellar populations, describe the stellar population properties 
equally (see also Groenewegen \& de Jong 1994).

\begin{figure*}
\includegraphics[clip=,width=0.8\textwidth]{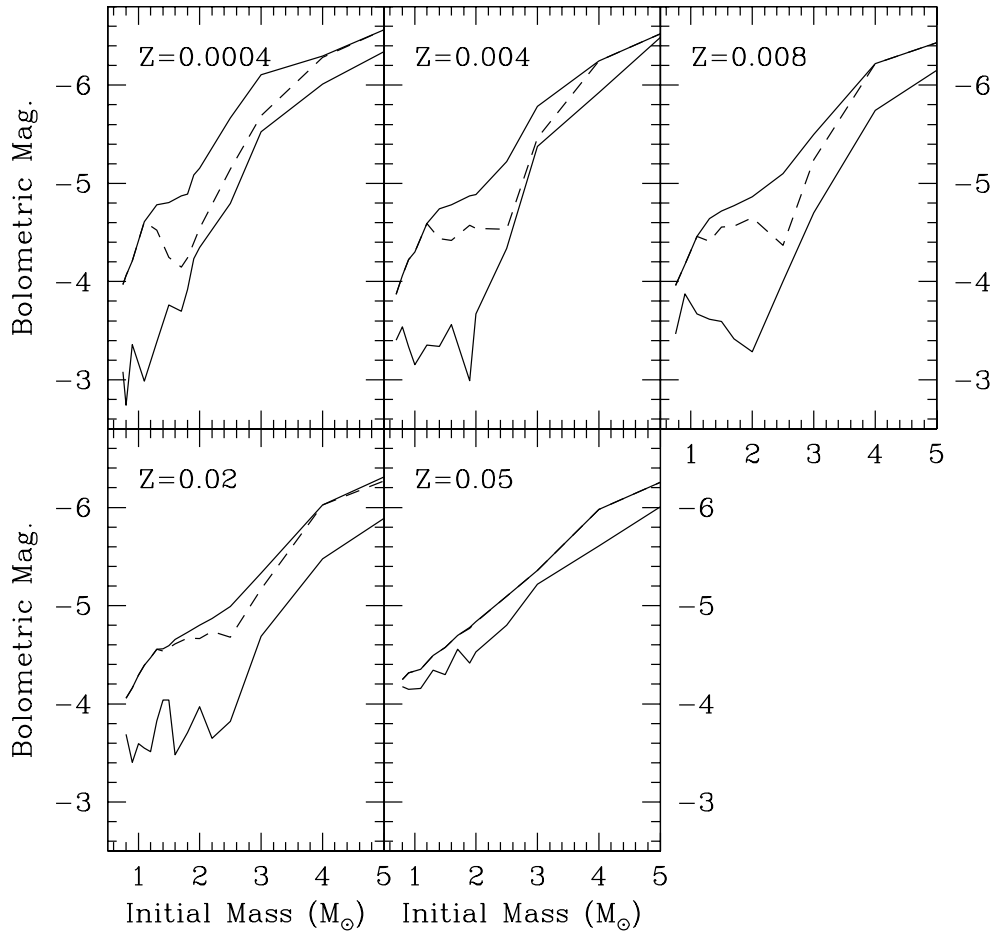}
\caption{Bolometric magnitudes versus initial mass for three characteristic
stages of the AGB phase, namely the start of the TP-AGB (bottom line), the
transition between the oxygen-rich phase and the carbon-rich phase (dashed
line), and the termination of the AGB phase. The predictions are those of 
our standard stellar model adopted in this work ($\alpha\,=\,2$, $\eta\,=\,0.1$ 
with Bl\"ocker's mass loss prescription).
The initial masses are in solar units, the initial metallicity, Z, is 
indicated on each panel.}
\label{grid_Z}
\end{figure*}

\begin{figure}
\includegraphics[clip=,angle=0,width=8.cm,height=11.cm]{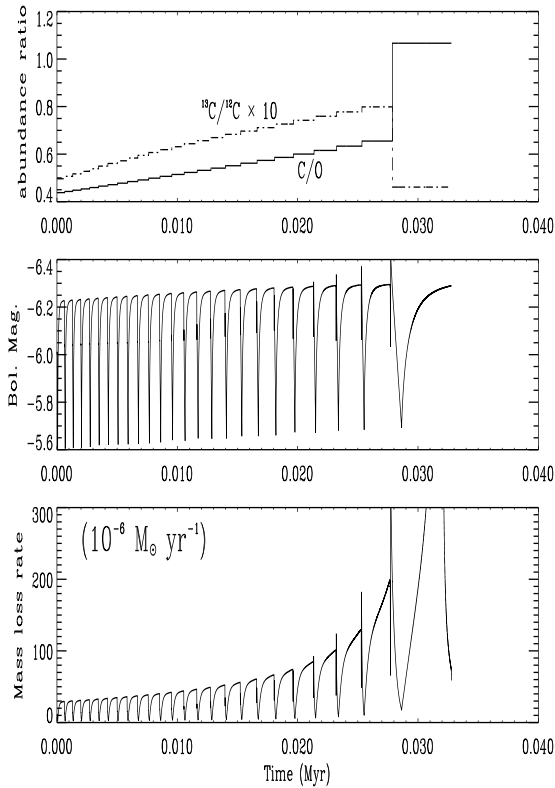}
\caption{AGB evolution of the $4\,M_{\odot}$ model with Z=0.0004. 
The top frame shows the 
$\rm{}^{13}{\kern-0.8pt}C$/$\rm{}^{12}{\kern-0.8pt}C$ and C/O ratios, 
the middle frame gives $M_{bol}$, and the lower frame the mass loss rate. }
\label{luminous_C_star}
\end{figure}

\subsubsection{The formation of carbon stars}
\label{Cstar_formation.sec}

At present, there seem to be two leading processes to form carbon 
stars: the first invokes dredge-up of carbon to the envelope during 
the AGB phase (Iben \& Renzini 1983), and the second involves binary 
mass transfer of carbon from a once more massive but now degenerate 
companion (Jorissen \& Mayor 1992). Carbon stars formed via the second 
channel fall below the red giant branch tip in the HR diagram. 
Those stars are observed without the radioactive s-process 
element technetium that specifically identifies stars on the TP-AGB 
(van Eck \& Jorissen 1999). They can be separated statistically from 
the intrinsic technetium-rich stars, as  the latter are on average cooler 
and brighter. Rebeirot et al. (1995) have surveyed the SMC for carbon 
stars. They found that faint carbon stars (M$_{bol}\,\le\,-3.2$) 
contribute less than a few percent, by number, to the total number of 
carbon stars (e.g. 4-5\%). In the rest of the paper carbon stars formed 
via binarity channel will not be counted. 
We consider only carbon stars formed via the third dredge-up channel. 
The formation of the evolution of carbon stars are investigated using
the model described in the previous subsections.

In Fig. \ref{grid_Z} we present the magnitudes at which stars with 
different initial masses leave the E-AGB phase and enter into the 
TP-AGB phase, magnitudes at which they become carbon stars, and 
finally magnitudes at which they leave the AGB phase and become 
Post-AGB stars (see also Iben 1981, Marigo et al. 1999). For each 
metallicity of the grid, the evolutionary tracks associate stellar 
ages with the transition magnitudes. 
%
%
As expected with the adopted mass loss prescription, the bolometric 
magnitude reached at the end of the TP-AGB phase depends on the 
metallicity as well as on the initial mass of the star; the lower 
the metallicity, the brighter the termination magnitude of the TP-AGB. 
In addition, the lower the metallicity, the larger is the fraction 
of the TP-AGB lifetime spent as a carbon star; the evolution of this 
fraction as a function of the initial stellar mass peaks around 
2\,-\,2.5\,M$_{\odot}$ independently of the initial metallicity. 
This mass range appears to be the privileged mass range for the formation 
of carbon stars at all metallicities (see Mouhcine \& Lan\c{c}on 2002
for more information). 

Fig. \ref{luminous_C_star} shows the evolution of bolometric magnitude, 
mass loss rate, C/O and $\rm{}^{13}{\kern-0.8pt}C$/$\rm{}^{12}{\kern-0.8pt}C$ 
of a [4\,M$_{\odot}$, Z=Z$_{\odot}$/50] model. The effect of HBB is clear: 
as the evolution proceeds, each dredge-up event brings about an increase of 
the C/O ratio
(with C/O\,=\,\-[n($\rm{}^{13}{\kern-0.8pt}C$)+n($\rm{}^{12}{\kern-0.8pt}C$)]/
[n($\rm{}^{16}{\kern-0.8pt}O$)+n($\rm{}^{17}{\kern-0.8pt}O$)], 
$\rm{}^{18}{\kern-0.8pt}O$ has a very small contribution). 
But as long as the envelope mass is high enough to maintain a hot temperature 
at the base of the envelope, the CNO cycle transforms dredged-up 
$\rm{}^{12}{\kern-0.8pt}C$ into $\rm{}^{13}{\kern-0.8pt}C$ and 
$\rm{}^{14}{\kern-0.8pt}N$. This increases the 
$\rm{}^{13}{\kern-0.8pt}C$/$\rm{}^{12}{\kern-0.8pt}C$ ratio 
and keeps the C/O ratio below unity. 
At the last thermal pulse, hot bottom burning has ceased because the large 
mass loss rate has dramatically reduced the mass of the envelope. The C/O 
ratio climbs abruptly. The model passes through C/O=1 and continues up to 
$\sim\,1.1$ at the time when the star ends the TP-AGB phase, reaching a 
bolometric magnitude of M$_{bol}\,=\,-6.3$.

From the initial mass-bolometric magnitude diagrams of Fig. \ref{grid_Z}, 
one can already deduce very interesting features concerning the evolution 
of populations of intermediate mass stars. The youngest AGB populations, 
observed in stellar populations with main sequence turn-off masses 
M$_{TO} \sim 5$\,M$_{\odot}$,
will be oxygen rich. As time goes by and the turn-off mass decreases, 
carbon stars will occupy increasing fractions of the TP-AGB, both in terms
of their luminosity range in the HR diagram and in terms of numbers.
A maximum is reached for M$_{TO}$\,=\,2-2.5\,M$_{\odot}$. The fraction
of the TP-AGB lifetime a star spends as a carbon rich object is largest 
for these initial masses (Mouhcine \& Lan\c{c}on 2002). Subsequently,
when the turn-off mass further decreases, this fraction shrinks
and the relative numbers of carbon stars drops.

The diagrams of Fig.\,\ref{grid_Z} display a strong sensitivity to 
metallicity. For the same initial mass, the fraction of the TP-AGB 
lifetime spent as a carbon star increases when the metallicity 
decreases. The following processes contribute to this result:  
(i) less dredge-up of carbon rich material from the core is needed 
to reach C/O\,$\geq$\,1 in metal-poor stars; 
(ii) at lower metallicities, the core mass at the beginning of the 
TP-AGB phase is larger, leading to earlier third dredge-up events; 
and (iii) the interpulse period is longer at lower metallicities, 
leading to more violent dredge-up events (see WG98 for more details); 
the He-shell accretes material for a longer time, and therefore 
more carbon-rich material is dredged-up to the envelope at each 
thermal pulse, while the envelope mass does not change significantly 
in comparison to the case where the metallicity effect on the interpulse 
period is neglected. 
%
%

\subsection{Chemical Evolution}
\label{chemevol.sec}

While star clusters may be represented with SSP models, the Local Group 
galaxies had more continuous SF histories. A self-consistent modeling 
of the carbon star populations of such systems must be able to account 
for chemical evolution and for the resulting coexistence of stellar 
generations with different initial metallicities and ages. Carbon stars 
are present and dominate AGB stellar populations over an age range of 
a few Gyr, during which the chemical evolution may be efficient. 
In this subsection, we will present the chemical evolution model used 
in this work.

The formation of stars, the re-ejection of gas, and the evolution of 
the gas mass evolve according to Tinsley's standard equation (1980) 
reducing the model to a set of few parameters that govern the evolution. 
The total gas mass evolves according to:    
\begin{equation}
\frac{dM_{gas}}{dt}=-\psi(t)+R(t)+f(t)
\end{equation}
where $\psi(t)$ is the star formation rate (SFR) and $R(t)$ is the gas 
return rate to the interstellar medium. $R(t)$ takes into account the 
finite lifetime of the turn-off stars, as derived from the stellar 
tracks also used to produce the synthetic stellar populations. 
We assumed instantaneous mixing of the ejected gas as well as instantaneous 
cooling of the hot gas component. The term $f(t)$ is an additional gas 
accretion rate from the intergalactic medium. An exponentially decreasing 
infall rate is assumed (Lacey \& Fall 1985):
\begin{equation}
f(t)=M_{tot}\,\frac{\exp*(-t/\tau_{f})}{\tau_{f}}
\label{amr_eq}
\end{equation}
where $M_{tot}$ is the total mass at 20\,Gyr and $\tau_{f}\,=\,5\,$Gyr 
is the accretion timescale for the formation of the galaxy. We assume 
that the accreted gas has primordial abundances (Walker et al. 1991). 
No outflow is considered in the models.

The evolution of the abundance $X_i$ of element $i$ in the interstellar 
medium obeys: 

\begin{eqnarray}
\frac{dX_{i}M_{gas}}{dt}&=&-X_{i}\psi(t)+R_{i,1}(t)+f_{SNIa}R_{i,2}(t)+\nonumber \\
             &&  (1-f_{SNIa})R_{i,3}(t)+R_{i,4}(t)+X_{i,p}f(t)
\end{eqnarray}
Here, $X_{i,p}$ is the abundance in the infalling gas, which is taken to be 
primordial. $R_{i,1(t)}$ and  $R_{i,2}(t)$ are described by the following 
equations:

\begin{equation}
R_{i,1}(t)=\int_{M_{low}}^{M_{Bmin}} \psi(t-\tau(m))\,\phi(m)\,mp_{i}(m)\, {\rm d}m
\end{equation}
and

\begin{eqnarray}
R_{i,2}(t)=\int_{M_{Bmin}}^{M_{Bmax}} \phi(m)\left[
           \int_{\mu_{min}}^{1/2} f(\mu)\psi(t-\tau(m))\,{\rm d}\mu\right] \nonumber \\
	    mp_{i}(m) {\rm d}m
\end{eqnarray}
$mp_{i}(m)$ is the stellar yield of the chemical element $i$ ejected by 
star of initial mass $m$, and $\phi$(m) is the initial mass function 
(hereafter IMF). 
In the rest of the paper we adopt a power law IMF between the lower and 
upper cutoff masses M$_{low}=\,0.8\,M_{\odot}$ 
and M$_{up}=\,120\,M_{\odot}$ ($\phi(m)\propto m^{-\alpha}, \alpha\,=\,2.35$).
 
$R_{i,2}(t)$ describes the enrichment from Type Ia supernovae (SNe Ia). 
These SNe Ia are assumed to originate in binary systems in which at least one 
of the stars is a white dwarf. The infall of gas from the companion pushes the 
mass above the Chandrasekhar limit, triggering a deflagration with subsequent 
disruption of the star. SNe Ia should be considered since they contribute a large 
fraction of iron. $M_{Bmin}$ and $M_{Bmax}$ are the lower and upper total masses 
of the relevant binary systems. We assume that the total mass of these binaries 
follows the same IMF as single stars. The initial mass range of the SNe Ia 
progenitors is taken from Greggio \& Renzini (1983). 
$f(\mu)$ is the distribution function of binary system mass ratios, defined as
the ratio between the secondary mass and the total mass $m$ of the system.
$R_{i,1}(t)$, $R_{i,3}(t)$ and $R_{i,4}(t)$ describe, respectively, the enrichment 
due to single stars with initial masses in different 
regions of the mass spectrum: $R_{i,1}(t)$ for stars with M$_{init}\,\le\,M_{Bmin}$, 
$R_{i,3}(t)$ for stars with M$_{init}\,\in\,[M_{Bmin}, M_{Bmax}$], and $R_{i,4}(t)$ 
for stars with M$_{init}\,\ge\,$M$_{Bmax}$.
$f_{SNIa}$ is the fraction of stars assumed to be in binary systems close 
enough to finally 
lead to SNe Ia events. A value of f$_{SNIa}\,\approx\,0.1$, used in our 
calculations, has been derived from the observed SNe I and II rates in 
our Galaxy (Evans et al. 1989).

The most important inputs into a chemical evolution calculation are the 
stellar yields. Three different sources of chemical elements must be 
considered: the first are the massive stars ($M\,>\,8-10\,M_{\odot}$) 
which evolve into Type II supernovae, the second are the intermediate 
mass stars ($0.8\,M_{\odot}\,<\,M\,<\,6\,M_{\odot}$) which evolve into 
planetary nebulae, and the third are the intermediate mass stars in close 
binary systems which become Type Ia supernovae.

For massive stars, we use the recent metallicity dependent stellar yields 
published by Portinari et al. (1998). They are based on the evolutionary 
tracks of the Padova group, and take into account both the material ejected 
by stellar winds and the explosive nucleosynthesis in SNe of Woosley 
\& Weaver (1995). For intermediate
mass stars, we use the synthetic evolution model discussed above to derive 
new metallicity dependent stellar yields (Mouhcine, in preparation). 
The yields for SNe Ia are taken from the improved $Z=Z_{\odot}$ W7 model 
presented in Thielemann et al. (1993). The mass of iron 
$\rm{}^{56}{\kern-0.8pt}Fe$ ejected is $0.63\,M_{\odot}$, with no remnant. 
This yield is assumed to be metallicity-independent, 
which is a reasonable assumption based upon the similarity of Thielemann 
et al's $Z=Z_{\odot}$ and $Z=0$ models. 

\begin{figure}
\includegraphics[clip=,angle=0,width=0.5\textwidth,height=9.cm]{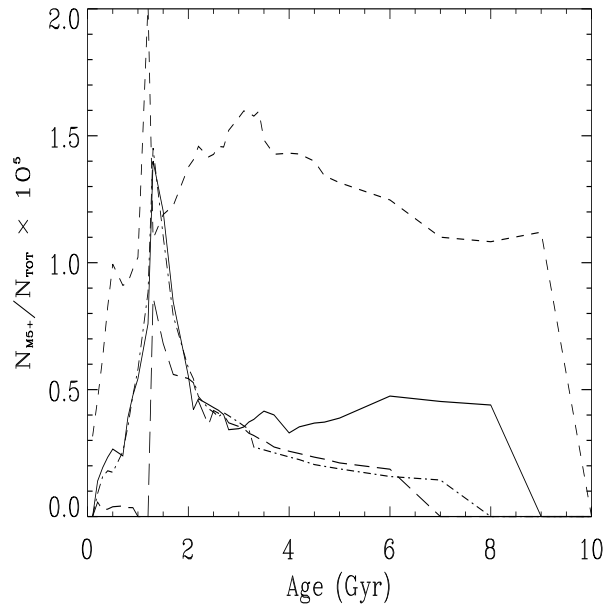}
\caption{Evolution of the fraction of late-type M stars as a function of age 
for single stellar population for Z=0.02 (dashed line), Z=0.008 (solid line),
Z=0.004 (long-dashed line), and Z=0.0004 (dot-dashed line). The effect of the
metallicity on the late-type M star efficiency is obvious; 
the lower the metallicity, the harder is the formation of late-type M stars, 
and the lower is their contribution to the total star number.}
\label{Mstar}
\end{figure}

\section{Statistics of AGB stars in a single stellar population}
\label{SSPs.sec}

In this section we discuss the model predictions regarding the statistics 
of intermediate mass stars in SSPs. The formation and the evolution of AGB 
stars is investigated using the model described in the previous section.

The observational techniques used to recognize AGB stars from other 
late type stars in external galaxies, and to classify them as oxygen-rich 
or carbon-rich, are based on colour criteria. 
We define carbon stars as stars evolving along the TP-AGB phase with 
C/O\,$>$\,1. 
The narrow band filter system mentioned in Sect.\,\ref{intro.sec} allows 
an empirical separation that agrees with this definition.
The formal definition of the late type oxygen-rich stars requires special 
understanding. 

For comparison with the data summarized in Section 2, our predictions 
regarding late type M stars must comply with the selection criteria currently 
applied by observers rather than correspond with a more theoretical definition 
of oxygen-rich TP-AGB stars. When only photometric data exists, stars with 
spectral type equal to M5 or later (M5+ hereafter), can in principle be defined 
as oxygen-rich stars redder (cooler) than a given colour (effective temperature), 
assuming a correspondence between colour (effective temperature) and spectral 
type. Richer et al. (1985) and Pritchet et al. (1987) defined M5+ stars as
stars with V-I\,$>$\,2. Albert et al. (2000) adopted R-I\,$>$\,0.9 as selection 
criterion. Measurements on solar neighborhood spectra show a good correspondance 
between these two criteria (see Fig. 1 of Lan\c{c}on \& Mouhcine, 2002)
We have to mention that those colour criteria are 
based on the earlier observations of LMC M stars of Blanco et al. (1980).  
In order to be able to compare our predictions with the observational 
findings, we adopt a colour criterion similarly to the observational 
procedures. We thus define ``M5+ stars" as oxygen rich stars with 
V-I\,$>$\,2. independently of the stellar population metallicity 
(Richer et al. 1985), and with an initial mass in the range of the mass 
spectrum bound to evolve through the AGB evolutionary phase. To project 
this colour criterion into the theoretical HR diagram, we use the 
metallicity dependent colour-effective temperature relation of Bessell 
et al. (1989).

For a single burst population of age $t$ and of total mass $M_{TOT}$, 
the total number of stars is
\begin{equation}
N_{SSP}(t)=M_{TOT}\,\int^{M(t)}_{0} \frac{\Phi(m)}{m}\,{\rm d}m,
\end{equation}
where $M(t)$ is the initial mass of the stars with lifetime $t$.
At an age $t$, stars of initial mass M$_{init}$\,=\,M$_{TO}(t)$ 
leave the main sequence (MS), while stars with 
M$_{init}$\,=\,M$_{TO}(t-T_{i})>$\,M$_{TO}(t)$ leave the post-MS 
evolutionary phase $i$. The number of stars between the main 
sequence and the end of phase $i$ is:
\begin{equation}
N_{i} \,=\,M_{TOT}\,\int_{M_{TO}(t)}^{M_{TO}(t-T_{i})} 
      \frac{\Phi(m)}{m}\,dm
\end{equation}
Using the approximations that: (i) in a SSP the post-main sequence 
stars occupy a very limited interval of initial masses, (ii) the 
typical lifetimes of different post-main sequence phases are very 
short in comparison with the MS lifetime of the turn-off mass, we 
can write (see e.g. Renzini \& Buzzoni 1986, Girardi \& Bertelli 1998): 
\begin{equation}
N_{i}\,\approx\,M_{TOT}b(t)T_{i}
\label{Ni}
\end{equation}
where 
$b(t)\,=\,\phi\left[M_{TO}(t)\right]/M_{TO}(t)\,
\left|dt/dM\right|^{-1}$ is the post-main sequence star production rate, 
and $T_{i}$ is the lifetime of a turn-off star between the turn-off and 
the end of phase $i$. Even though this calculation involves simplistic 
approximations, it provides us with a tool for a qualitative understanding 
of the evolution of the N$_{C}$/N$_{M5+}$ ratio, since it can be used to 
relate the number ratio directly to the ratio of the durations of the 
carbon-rich phase ($T_c$) and the late M-type phase ($T_{M5+}$).

\begin{equation}
\frac{N_{c}}{N_{M5+}}(t)\,
\approx\,\left[\frac{T_{c}}{T_{M5+}}\right]_{M_{TO}(t)}
\end{equation}

\begin{figure}
\includegraphics[clip=,angle=90,width=0.5\textwidth,height=9.cm]{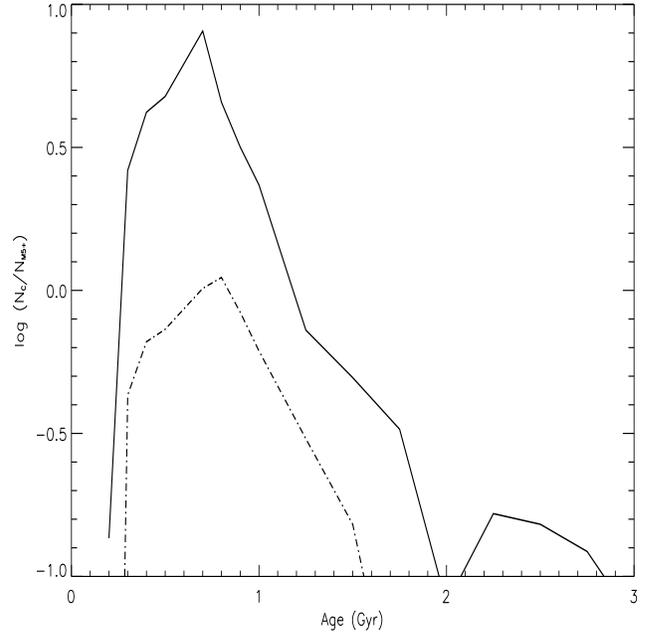}
\caption{Temporal evolution of the carbon stars to M5+ stars ratio vs. metallicity 
for instantaneous burst scenario for Z=0.008 (continuous line) and Z=0.02 
(dotted-dashed line). The effect of the initial stellar population abundances 
is clear; the lower the metallicity, the higher is the ratio.}
\label{stat_Z}
\end{figure}

In Figure \ref{Mstar} we show the time evolution of the fractional 
number of M5+ stars among all stars,  
N$_{M5+}$/N$_{TOT}$, in stellar populations with metallicities between
Z/Z$_{\odot}$\,=\,1/50 and Z/Z$_{\odot}$\,=\,1. The models evolve from 
an instantaneous burst with a Salpeter IMF. To compute N$_{TOT}$ we  
consider only stars that contribute to the population light budget 
(i.e. the stellar remnants are not considered). The models evolve from 
an instantaneous burst with the Salpeter IMF specified in 
Sect.\,\ref{chemevol.sec}. We predict a rapid 
increase of the fraction of M5+ stars with time, with a maximum at 
about 1\,Gyr because the duration of the TP-AGB reaches a maximum 
for stars with these lifetimes.
After the maximum, the number fraction of M5+ stars evolves and decreases 
slowly. It goes to zero when the coolest temperature on the isochrone 
is higher than that of an M5+ star.
The striking features shown in Fig.\,\ref{Mstar} are that the maximum 
N$_{M5+}$/N$_{TOT}$ ratio {\em and} the range of ages over which M5+ 
stars are numerous decrease rapidly with metallicity.

Figure \ref{stat_Z} displays the evolution of  N$_{C}$/N$_{M5+}$, 
for solar and LMC metallicities. As expected from 
Fig.\,\ref{grid_Z}, carbon stars are predominant at low metallicities.
In addition, at a given metallicity, the distribution of N$_{C}$/N$_{M5+}$  
evolves with time. The ratio rises rapidly up to a maximum 
just before 1\,Gyr of age, and decreases for older systems. This temporal 
evolution of  N$_{C}$/N$_{M5+}$ is directly related to changes in the
duration of the carbon rich phase, $T_{c}$, 
relative to the total lifetime in the TP-AGB phase, $T_{\rm TP-AGB}$,
as the turn-off mass of the stellar population decreases.  
This relative duration is largest for stars with initial masses in the 
range of $2\,-\,2.5\,M_{\odot}$, which have a main sequence lifetime  
close to 1\,Gyr.

In summary, metallicity has two main consequences on the populations 
of late-type stars. The number proportion of M5+ stars decreases 
intrinsically with metallicity because the giant branch becomes bluer 
with decreasing metallicity. At a given age after a starburst, 
N$_{C}$/N$_{M5+}$ is a strong function of metallicity. One relatively
metallicity-independent property is that the maximum N$_{C}$/N$_{M5+}$ 
occurs when the turn-off mass is of 2 -- 2.5\,M$_{\odot}$, because stars 
with these initial masses have the longest carbon rich phase independently
of the stellar metallicity.
\medskip 

As a step towards the study of the complex stellar populations of 
galaxies, we briefly look at the late type stars of models with a 
constant SFR and a fixed metallicity.
The proportion of M5+ stars is given by the averages
of the curves in Fig.\,\ref{Mstar} over the accumulated range of ages. 
The fractional number of M5+ stars increases over the first
1--2\,Gyr, then stabilises for many billions of years. It 
is reduced progressively only at very old ages.
As a result of the larger contrast
in Fig.\,\ref{stat_Z} (note the logarithmic scale), N$_{C}$/N$_{M5+}$ 
keeps a strong dependence on time even when the SFR is constant. 
Because N$_{C}$/N$_{M5+}$ rises extremely rapidly, stars 
of the first generation will dominate the 
number of carbon stars during the first Gyr of evolution.
The population of carbon stars is essentially stationary after
the first $\sim 1-1.5$\,Gyr. Carbon stars with lower initial masses 
that remain to be born represent only a tiny fraction of the carbon
stars that already appear and die in a stationary fashion. As a 
result, N$_{C}$/N$_{M5+}$ slowly decreases by dilution after the
first few billions of years. The maximal and the quasi-stationary
values reached by N$_{C}$/N$_{M5+}$ both increase with decreasing Z. 
It is interesting to mention that the difference between the 
quasi-stationary N$_{C}$/N$_{M5+}$ of a constantly star forming 
galaxy model and the maximum N$_{C}$/N$_{M5+}$ reached after a 
burst is larger at lower metallicities. Indeed, in metal poor 
environments the range of initial masses capable of producing 
carbon stars is broader than at high Z; the carbon star episode 
following a burst is more stretched in time, and the contrast
with a constant star formation situation is less pronounced.

\section{Effects of mass loss on carbon star statistics.}

The majority of observational techniques applied to search for 
carbon stars operate in the optical. 
Those techniques are biased towards unobscured carbon stars and 
may be missing the majority of mass loosing stars. The transition 
from optically visible stars to purely infrared sources is 
related to the mass-loss which produces the circumstellar envelopes.
It is important for our purpose to know when this transition occurs, 
and how much time a TP-AGB star spends as a dust-obscured object. 
In this section we derive an estimate of the number fraction of 
carbon stars that may be missed in optical surveys. 

To estimate the lifetime fraction that a carbon star spends as 
optically invisible, we assume that the stars is no longer observable 
when the optical depth at $1\mu$m, $\tau_{1\mu m}$, is higher than 
a given critical value ($\tau_{1\mu m}\ga\,\tau_{crit}$). 
To estimate the optical depth we have to construct dusty envelope 
models around late type AGB stars. To do so we assume (i) spherical 
symmetry, (ii) constant mass outflow and expansion velocity. 
Neglecting the inverse of the circumstellar shell outer radius with 
respect to the inverse of the grain condensation radius, which is 
not bad approximation as we restrict ourselves to the near-infrared
(Rowan-Robinson 1986), the optical depth as a function of wavelength 
is given by:
\begin{equation}
\tau_{\lambda} \propto\,\frac{\dot{M}\,\Psi}{R_c\,v_{d}}
                         \frac{Q_{ext}(\lambda,a)/a}{\rho_{gr}(a)} 
\end{equation}
where Q$_{ext}(\lambda$) is the sum of the absorption and scattering 
coefficients of the dust, 
$a$ the grain radius, $\rho_{gr}$ the grain density, R$_c$ the inner 
radius of the dust shell, called also the dust
condensation radius, $v_{d}$  the dust outflow velocity, and $\Psi$ 
the dust-to-gas ratio. The condensation radius is taken to vary as 
$R_c,\propto\,\left[T_{eff}/T{_c}\right]^{(4+\beta)/2}$ where the 
parameter $\beta$ gives the overall wavelength dependence of the 
grain properties, $Q(\lambda)\propto\lambda^{-\beta}$ (Jura 1983, 
Bedijn 1987), and equals $\sim\,1$ for amorphous carbon dust. Grain 
properties at $1\mu$m are taken from Suh (2000) for $a=0.1\mu$m 
grains, with a grain density of $\rho_{gr}\sim\,2.5\,$gr\,cm$^{-3}$. 
The condensation temperature of amorphous carbon grains is taken 
to be T$_{c}=1000\,K$ (Le Sidaner \& Le Bertre 1996). The dependence 
of both $\Psi$ and $v_{d}$ on the fundamental stellar parameters 
are taken from Habing et al. (1994) and Vassiliadis \& Wood (1993).

\begin{figure}
\includegraphics[clip=,angle=0,width=0.5\textwidth,height=9.cm]
{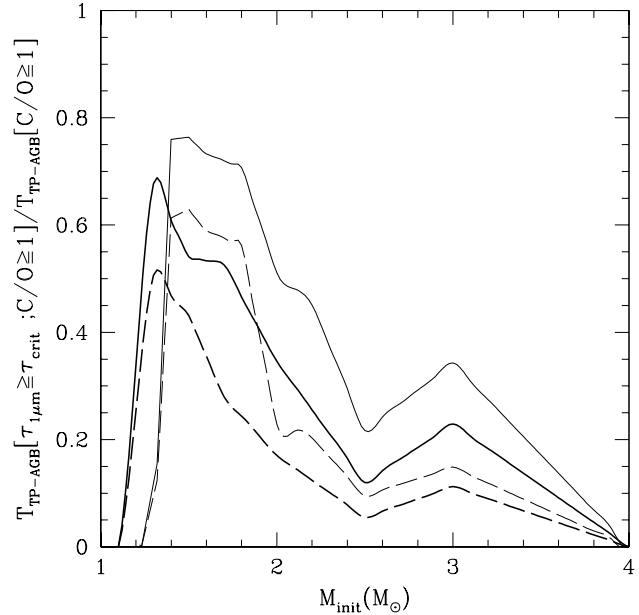}
\caption{Fraction of the total carbon rich lifetime spent as an 
optically invisible infrared source, as a function of initial mass, 
for Z=0.02 (thin lines) and Z=0.008 (thick lines) metallicities.
Stars are assumed to be no longer optically visible when the optical 
depth at $1\mu$m is larger than the unity (continuous lines) or than 
3 (dashed lines). The plot shows that higher the star metallicity, 
larger is the fraction of carbon-rich phase spent as being optically 
invisible. See text for more details. }
\label{Cdust}
\end{figure}

Figure \ref{Cdust} shows the evolution of the fraction of its 
carbon-rich lifetime that a star spends as an optically invisible object, 
as a function of initial mass for Z=0.02 and Z=0.008.
Two different choices of the critical optical depth ($\tau_{crit}=1,3$) 
are shown. This range of optical depths was chosen to reflect 
the uncertainties on different ingredients used to construct the 
envelope models, known in order of 1-3 (Jura, private communication).

The most striking feature, from the population synthesis point of view, 
that is coming out from this plot is the dip at initial masses of 
$\sim\,2.5\,$M$_{\odot}$. One has to keep in mind that those stars 
form the bulk of the carbon stars as shown in Sec.\,\ref{SSPs.sec}. 
These stars remain optically visible for most of their carbon rich 
phase. Their fundamental parameters are such that they trigger a 
superwind only very late in their evolutionary history. 
This is in agreement with the observational finding that the majority 
of carbon stars have low mass loss rates.
The plot shows also that decreasing initial stellar metallicity 
decreases the lifetime fraction that a carbon star spend being optically 
invisible, reflecting the extreme sensitivity of the optical depth 
to metallicity. This is of great importance for our purpose as carbon 
stars as found preferentially in metal-poor systems, and that the 
galaxy sample for which carbon star statistics are available in the 
literature is composed mainly by metal-poor galaxies. 

As a consequence, the numerical values in Fig.\,\ref{Cdust} at the
smallest and largest initial masses are very sensitive to choices that
enter our definition of the transition from optically visible to obscured
stars or from oxygen-rich to carbon-rich stars. In comparison, the results 
for intermediate initial masses are robust, because all relevant evolutionary
phases are of longer duration. 

Using Eq. \ref{Ni}, one should expect that, for a single stellar
population, the number fraction of optically invisible carbon stars 
among all carbon stars will behave as the ratio of the lifetimes in 
the corresponding phases as shown in Fig.\,\ref{Cdust} 
($N_{CIR}/N_{C}\approx\,[T_{CIR}/T_{C}]_{M_{TO}}$, where $M_{TO}$ is 
the current turn-off mass). Hence when the stellar population evolves, 
the fraction of dust-surrounded carbon stars decreases with a minimum 
at $\sim$1\,Gyr, when M$_{TO}\sim 2.5$M$_{\odot}$. For older stellar 
populations, the fraction of obscured carbon stars is expected to 
increase while, as discussed already, the whole population of carbon 
stars shrinks in importance.

We conclude that the contribution of obscured carbon stars to the total 
number of carbon stars is relatively small (10\,-20\,\% at $\sim\,1$Gyr) 
and that the global statistics of carbon stars will not be affected 
dramatically if one uses only the carbon stars found in optical surveys. 
In the rest of the paper we will calculate the evolution of the total 
number of carbon stars including optically visible and invisible sources.

\section{Statistics of AGB stars in complex stellar populations.}

Using the results of the previous sections, it is possible to give 
theoretical predictions for the evolution of various late-type stellar 
populations for stellar systems with complex SF histories.

The number of carbon stars and late type M stars for a stellar population 
with a star formation rate $\psi$(t) is given by:
\begin{equation}
N_{C}(t)=\!\! \int\!\!\!\!\int \psi(t-\tau(m))\frac{\phi_{AGB}(m)}{m} 
               \epsilon_C(m,\tau,Z) \, \rm{d}m\, \rm{d}\tau        
\end{equation}
\begin{equation}
N_{M}(t)= \!\! \int\!\!\!\!\int \psi(t-\tau(m))\frac{\phi_{AGB}(m)}{m} 
          \epsilon_M(m,\tau,Z) \, \rm{d}m\, \rm{d}\tau         
\end{equation}
where $\phi_{AGB}(M)$ is the IMF restricted to the mass range of stars 
that indeed evolve through an AGB phase. The adopted lower and higher 
initial masses of AGB progenitors depend slightly on the metallicity. 
$\epsilon_C(m,t,Z)$ and $\epsilon_M(m,t,Z)$ are derived from the 
evolution of the C/O abundance ratio in AGB star envelopes.
$\epsilon_C(m,t,Z)$ describes, for a given initial mass and 
initial metallicity, when and for how long a TP-AGB star is
carbon rich. $\epsilon_C(m,t,Z)$ is given by:
\begin{equation}
\epsilon_C(m,t,Z)=
\left\{
\begin{array}{rl}
         1  \qquad \mbox{if} \,\,\, t \geq f_C(m,Z)T(m,Z)\\
         0  \qquad \mbox{if} \,\,\, t  <  f_C(m,Z)T(m,Z)
\end{array}
\right.
\end{equation}
$T(m,Z)$ is the time that elapses between the moment a star of initial 
mass $m$ becomes and M5+ object and the moment it leaves the TP-AGB;
$f_C(m,Z)$ is the fraction of this time the star spends as a carbon 
star. Similarly, $\epsilon_M(m,t,Z) = 1 - \epsilon_C(m,t,Z)$ indicates 
which stars are to be counted as M5+ objects. 

As mentioned in Sect.\,\ref{chemevol.sec},
we restrict ourselves to one-zone galaxy models, assume ideal mixing, 
and start the chemical evolution from a homogeneous gas 
cloud of primordial composition. The global SFR is
assumed to be a function of the total gas content (Schmidt 1963), 
with the star formation efficiency as a free parameter. 
The efficiency is chosen such as to mimic the  SF
histories along the Hubble sequence, as the spectrophotometric
properties of this galaxy sequence  
is primarily dictated by the characteristic time scale for the 
star formation (e.g. Sandage 1986, Charlot \& Bruzual 1991, 
Fritze von-Alvensleben \& Gerhard 1994, Fioc \& Rocca-Volmerange 1997). 
\begin{equation}
\psi(t)\,=\,\frac{\rho_g^k(t)}{\tau_{*}}
\end{equation}
where $\tau_{*}$ is the star formation timescale and $\rho_g$ the 
available gas mass. 
We will use the timescales of Fioc \& Rocca-Volmerange (1997) here. 
As the Local Group galaxies are mostly dwarfs (irregulars or spheroidals) 
or late-type galaxies, we will restrict our calculations to timescales 
typical of Sa-type to Irr-type galaxies. 
As we are interested in the systematic behavior of carbon star statistics 
and are not focusing on any particular stellar system, we do not consider 
other forms of SFRs (bursting star formation for example). 

\begin{figure}
\includegraphics[clip=,angle=0,width=0.5\textwidth,height=9.cm]{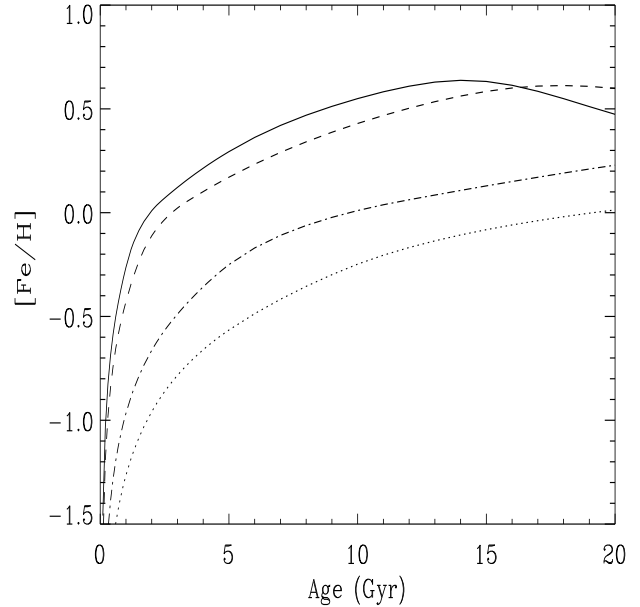}
\caption{Age-metallicity relation for the different star formation
histories discussed in this paper, assuming closed box galaxy models. 
[Fe/H] refers to the interstellar matter.
Different lines refer to galaxies with different SF timescales $\tau_{*}$.
The solid line represents Hubble type Sa, the dashed line type Sb,
the dot-dashed line type Sc and dotted line Irregular galaxies.
The longer the SF timescale, the slower the chemical evolution (see text).}
\label{AMR}
\end{figure}

Figure \ref{AMR} shows the age-metallicity relation for 
the closed box galaxy models of this paper, with Schmidt's 
exponent $k$ equal to unity. The metallicity rises rapidly 
at early times. The rate of rise in [Fe/H] is largest when
the SF timescale is short, i.e. with the 
SF scenarii appropriate for early type spiral galaxies.
These models also have the largest initial star formation
efficiency, and they are the first to reach their maximum 
star formation rate. The AMRs flatten for older systems,
mainly due to decreasing star formation rates. 
When star formation has been negligible for a long 
enough period of time, [Fe/H] may decrease as a result of 
the dilution of interstellar iron in iron-poor low mass star 
ejecta (the iron abundance is significantly determined by the 
contribution of SNe Ia). 
The chemical evolution in models with gas infall (Eq.\,\ref{amr_eq})
is qualitatively similar, except that it is slower
because the accreted gas is assumed to be metal-free.

\begin{figure*}
\includegraphics[clip=,angle=0,width=0.45\textwidth,height=11.cm]
{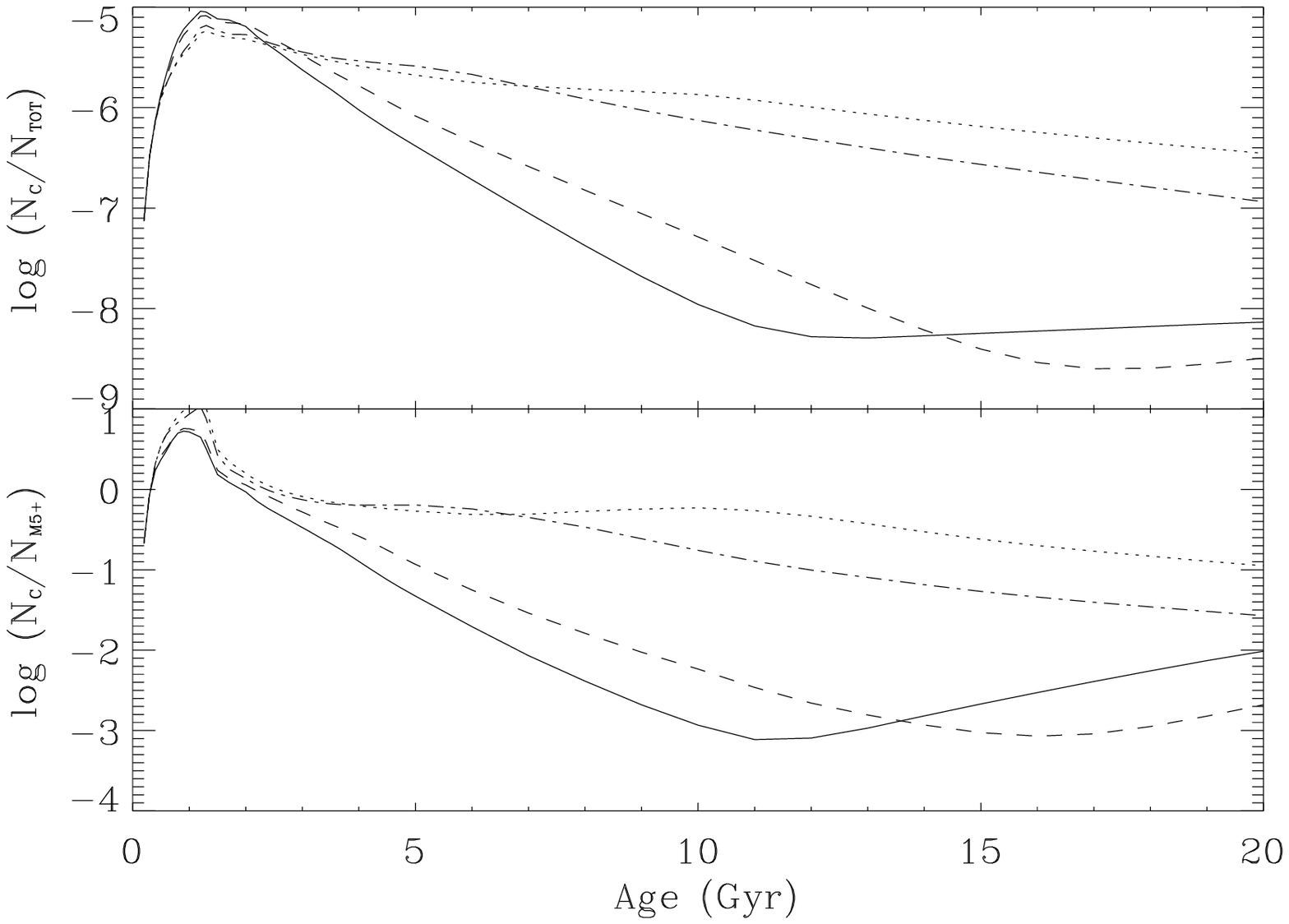}
\includegraphics[clip=,angle=0,width=0.45\textwidth,height=11.cm]
{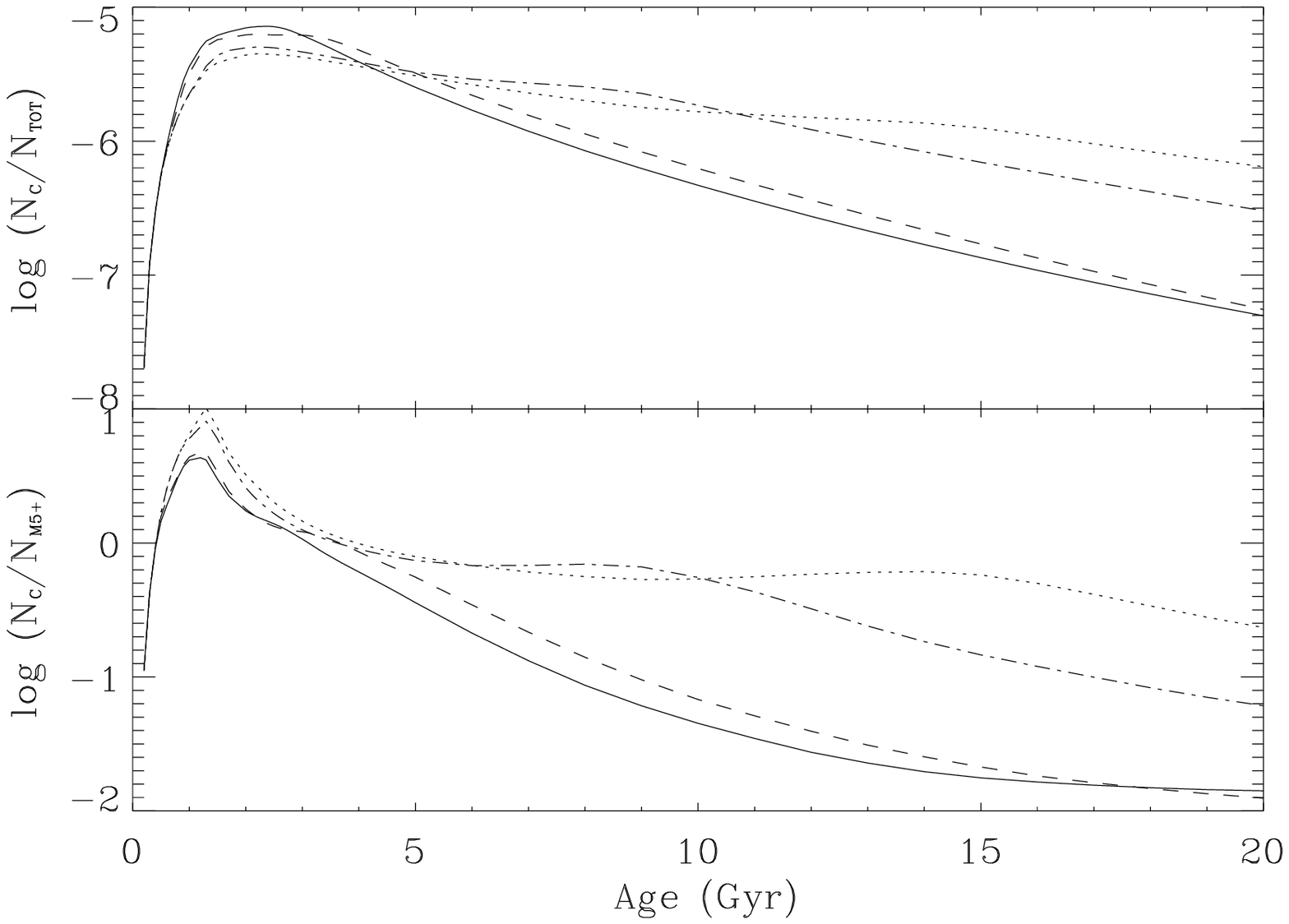}
\caption{Left: Evolution of the logarithmic number ratio of carbon rich to late 
M stars (bottom plot), and of the logarithmic number fraction of carbon rich 
among all stars (top plot). The lower the metallicity, the higher is the carbon 
star contribution to the total number of stars. The chemical evolution is computed 
with the closed box assumption. Right: Similar to the left panel but assuming that 
the stellar populations are formed by gas infall (see text)}
\label{number_Nc}
\end{figure*}

Figure \ref{number_Nc} shows the evolution of N$_{C}$/N$_{M5+}$ and 
of the relative number of carbon stars in the whole stellar population 
for the SF histories considered above, for both closed box and infall 
models, and $k=1$. The generic behavior of N$_{C}$/N$_{M5+}$
as a function of age can be understood as follows, in the light of
the discussion in Sect.\,\ref{SSPs.sec}. 
Regardless of the details of the star formation history process (i.e. the
SF timescale) or the galaxy formation (closed box or infall model), the 
ratio sharply increases and reaches its maximum at an age of $\sim$\,1\,Gyr.
When the massive carbon stars of the first generation of stars die, 
two types of carbon stars replace them. A second generation of 
carbon stars with the same high initial mass and a slightly higher 
metallicity appears, but also carbon stars with lower initial mass 
appear for the first time. The latter have the same metallicity as the 
more massive carbon stars that are dying. A longer carbon rich phase
is associated with their lower initial mass. The increase in the 
duration of the carbon rich phase is rapid compared to changes
in the SFR. As a consequence, those lower mass carbon stars will dominate, 
by number, the whole carbon star population. The relative
number of carbon stars continues to increase until stars with 
the longest carbon rich phase (i.e. initial masses of 2.5--2\,M$_{\odot}$
and main sequence lifetimes of the order of $1$\,Gyr; see Fig \ref{grid_Z}) 
evolve off the main sequence. Then, a quasi-stationary carbon star 
population is established: the birth and death rates of the predominant
carbon stars are balanced. Changes in the carbon star proportions then 
essentially reflect the evolution of the metallicity, with a delay 
of the order of $1$\,Gyr which is short compared to the evolutionary 
timescale of the SFR. N$_{C}$/N$_{M5+}$ decreases, tracing the metallicity 
increase with increasing age and with decreasing star formation timescale.

It is noteworthy that the maximum reached by the N$_{C}$/N$_{TOT}$ ratio 
increases with decreasing star formation timescale.
Indeed, the carbon stars of the first generations are compared to a total 
number of stars that will be smaller if star formation decreases rapidly. 
In contrast, the maximum reached by the N$_{C}$/N$_{M5+}$ ratio decreases 
with decreasing SF timescales. This essentially reflects the evolution of 
metallicity, as the dominant carbon and M5+ stars are of similar age.

In summary, the figures show the predominance of carbon stars relative 
to late type M stars and the high carbon star formation efficiency 
for metal-poor systems corresponding to stellar systems with 
low star formation rate. 
Models calculated assuming gas infall present the same qualitative 
behavior but with scaled-up N$_{C}$/N$_{M5+}$ and N$_{C}$/N$_{TOT}$ 
ratios due to the slower chemical evolution.

In the following subsections, we will compare our number count models 
combining the evolutionary population synthesis models and the chemical 
evolution models with the observed properties described in 
Sec.\,\ref{Obs.sec}.

\subsection{The evolution of N$_{C}$/N$_{M5+}$ as a function of metallicity}

\begin{figure*}
\includegraphics[clip=,angle=0,width=17.cm,height=10.cm]{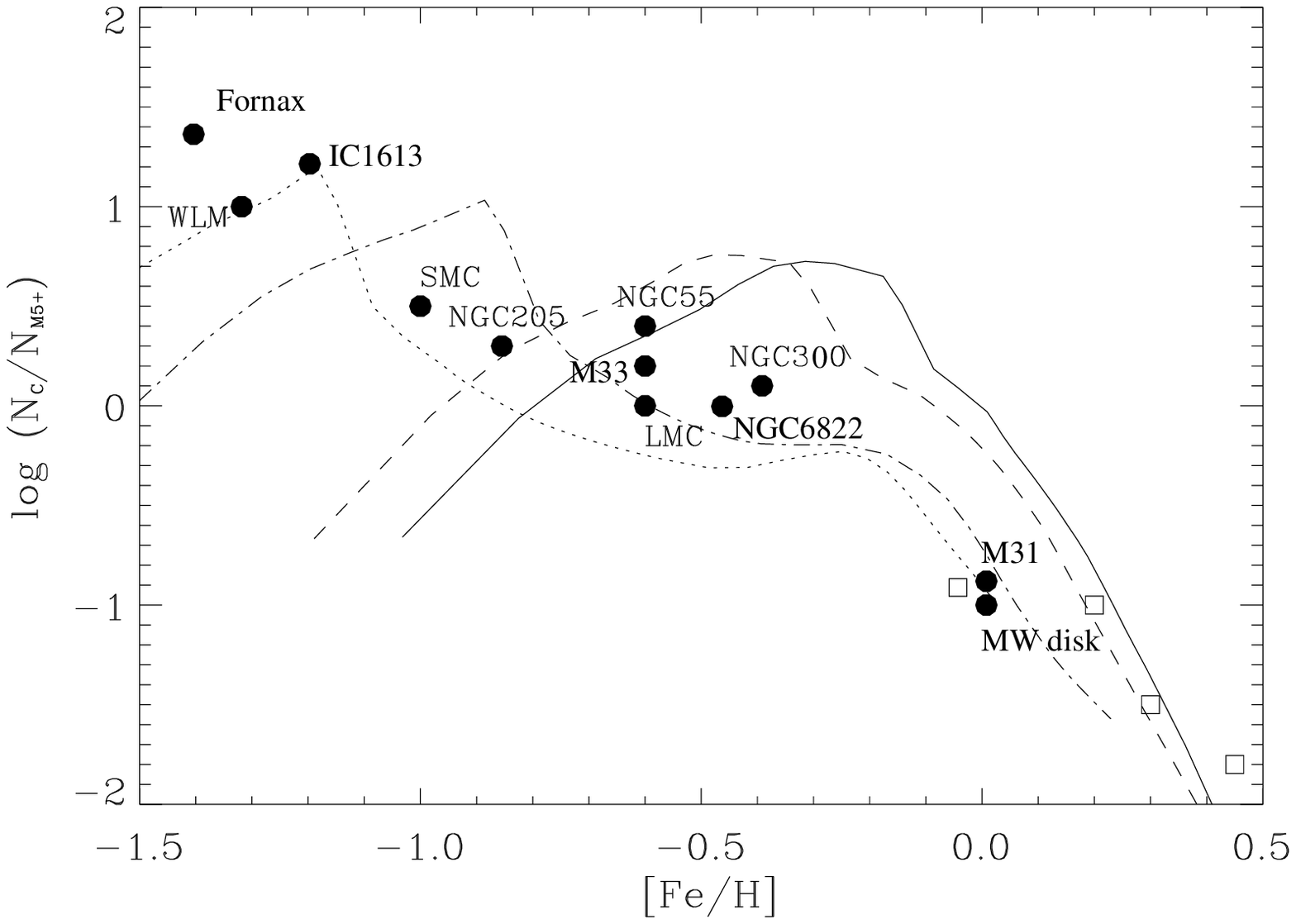}
\caption{The logarithmic ratio of the number carbon stars to the
number of late M-type stars, as a function of the metallicity of 
the host galaxy (gas phase). The chemical evolution assumed here is a 
closed box. Different lines refer to different star formation histories.
For stellar populations older than about 1\,Gyr,
all the models merge into a common sequence. 
Also shown are the observational data for the Local Group galaxies 
from the recent compilation of Groenewegen 1999.}
\label{Nc_Nm_fetoh}
\end{figure*}

\begin{figure*}
\includegraphics[clip=,angle=0,width=17.cm,height=10.cm]{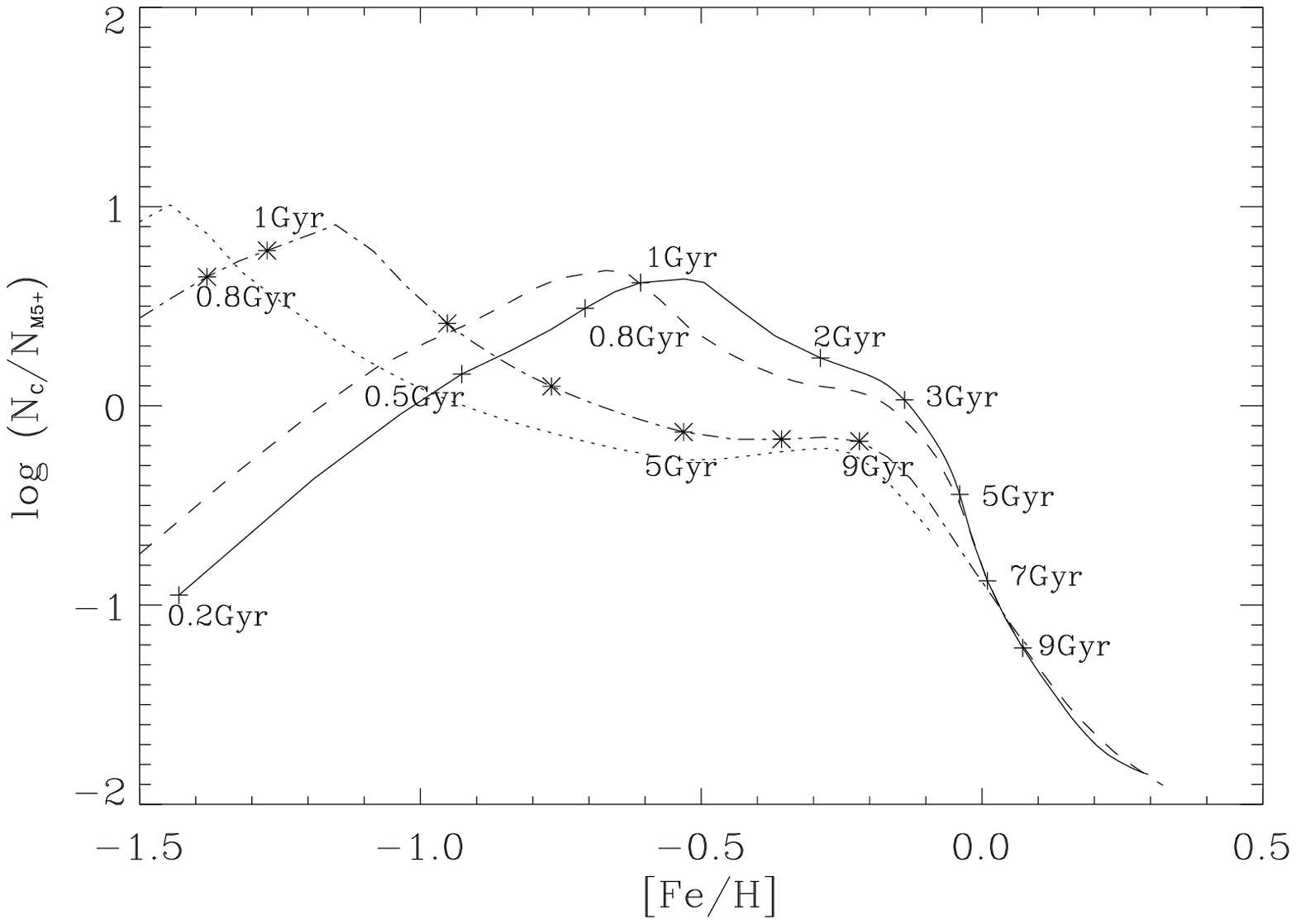}
\caption{Same as Fig. \ref{Nc_Nm_fetoh}, but assuming that the stellar
population is formed by gas infall (see text). Also show are a sample of 
ages of the considered stellar population for different star formation 
histories.}
\label{Nc_Nm_fetoh_infall}
\end{figure*}

By combining the temporal evolution of N$_{C}$/N$_{M5+}$ and of the 
metallicity, we can derive the evolution of N$_{C}$/N$_{M5+}$ as function of 
[Fe/H]. Figures \ref{Nc_Nm_fetoh} and \ref{Nc_Nm_fetoh_infall}, respectively,
show the resulting predictions in the case of closed box and infall models,
with Schmidt's exponent $k=1$. For direct comparison, the observational data 
for galaxies with good carbon star statistics are over-plotted on 
Fig.\,\ref{Nc_Nm_fetoh}. N$_{C}$/N$_{M5+}$ ratios are taken from Groenewegen 
(1999), apart for NGC\,8266 taken from Letarte et al. (2002) and IC\,1613 
estimated from Albert et al (2000) data set.  
[Fe/H] represents the best available estimate of the interstellar medium
metallicity. Very good agreement is found. The predicted sequence in the 
N$_{C}$/N$_{M5+}$ vs. [Fe/H] plane exhibits the same trend as the observed 
one. Metal-poor stellar systems have a larger number of carbon stars relative 
to number of late type M stars. 
The theoretical curves delineate quite well the boundaries of the locus of
the observational data points. 
To give the reader an idea about the time evolution effect on the considered 
diagram, some reference ages are indicated in Fig.\,\ref{Nc_Nm_fetoh_infall}.

For stellar populations younger than 0.8-1\,Gyr, we observe a dramatic 
evolution of N$_{C}$/N$_{M5+}$ as a function of metallicity, 
It must be kept in mind that the dominant
carbon star populations in those systems have a narrow range of (low) 
stellar metallicities. The first generation of stars with initial mass 
in the privileged range of 2--2.5\,M$_{\odot}$ produces carbon stars 
after about 1\,Gyr; carbon stars with these progenitor masses 
remain predominant. 
The star formation timescale determines how rapidly the 
interstellar medium is enriched at early times.
For older stellar systems, carbon stars form at a
quasi-stationary rate, and N$_{C}$/N$_{M5+}$ essentially traces 
metallicity. Its quasi-monotonic evolution with metallicity is a 
natural consequence of the evolution of carbon star properties as a
function of metallicity. The number ratio expresses the metallicity 
dependance of the ratio between the durations of the carbon rich phase and 
of the M5+ phase, weighted by the IMF. The models for old populations
form a well-defined sequence in the N$_{C}$/N$_{M5+}$ vs. [Fe/H] diagram.
The locus of the models depends very little on the assumed SF history.
The gas infall process has no major effects on this locus either. 
Considering infall will only slow down the chemical evolution of the stellar 
population, and consequently the stationary carbon star formation will 
take place at lower metallicity, with a higher N$_{C}$/N$_{M5+}$ ratio as 
for the closed-box models.
This is consistent with the observational result pointed out by 
Brewer et al. (1995), i.e. that the N$_{C}$/N$_{M5+}$ ratio is determined 
by the local metallicity rather than by galactic morphological 
(or spectroscopic) type. 

The evolution of the statistics of carbon star populations as a function 
of metallicity shows a very small sensitivity to the exponent of the star 
formation law, which affects the shape of the SFR. This is consistent with 
the result discussed above: the global properties of carbon star populations 
are determined mainly by metallicity rather than by the shape of the SF 
history.

In the rest of the paper we restrict ourselves to $k$\,=\,1.

\subsection{The mean carbon star luminosity as a function of metallicity}

As discussed in Sect.\,\ref{Obs.sec}, the second observed property
of carbon star populations is their mean bolometric luminosity.  
We use the following definition:
\begin{equation}
<\!M_{Bol,C}\!> = -2.5\,\log\,\left(\frac{\sum_{i=1}^{N_{C}}L_{i,C}}{N_{C}}\right)+4.74
\end{equation}
where $\sum_{i=1}^{N_{C}}L_{i,C}$ is the sum of the luminosity contributions 
of all carbon stars, and N$_{C}$ is the total number of carbon stars.

Metallicity has a significant effect on the time evolution of the
luminosity of the brightest TP-AGB stars of single stellar populations.
Indeed, metallicity affects the mass loss rate, and hence the initial-final
mass relation: the lower the metallicity, the lower is the mass loss rate,
and the higher are the final core mass and bolometric luminosity
(Mouhcine \& Lan\c{c}on 2002, Zijlstra 1998). At a given turn-off
mass, the brightest carbon stars of a metal-rich SSP will be fainter
than those of a metal-poor SSP.

\begin{figure*}
\includegraphics[clip=,angle=0,width=0.45\textwidth,height=9.cm]
{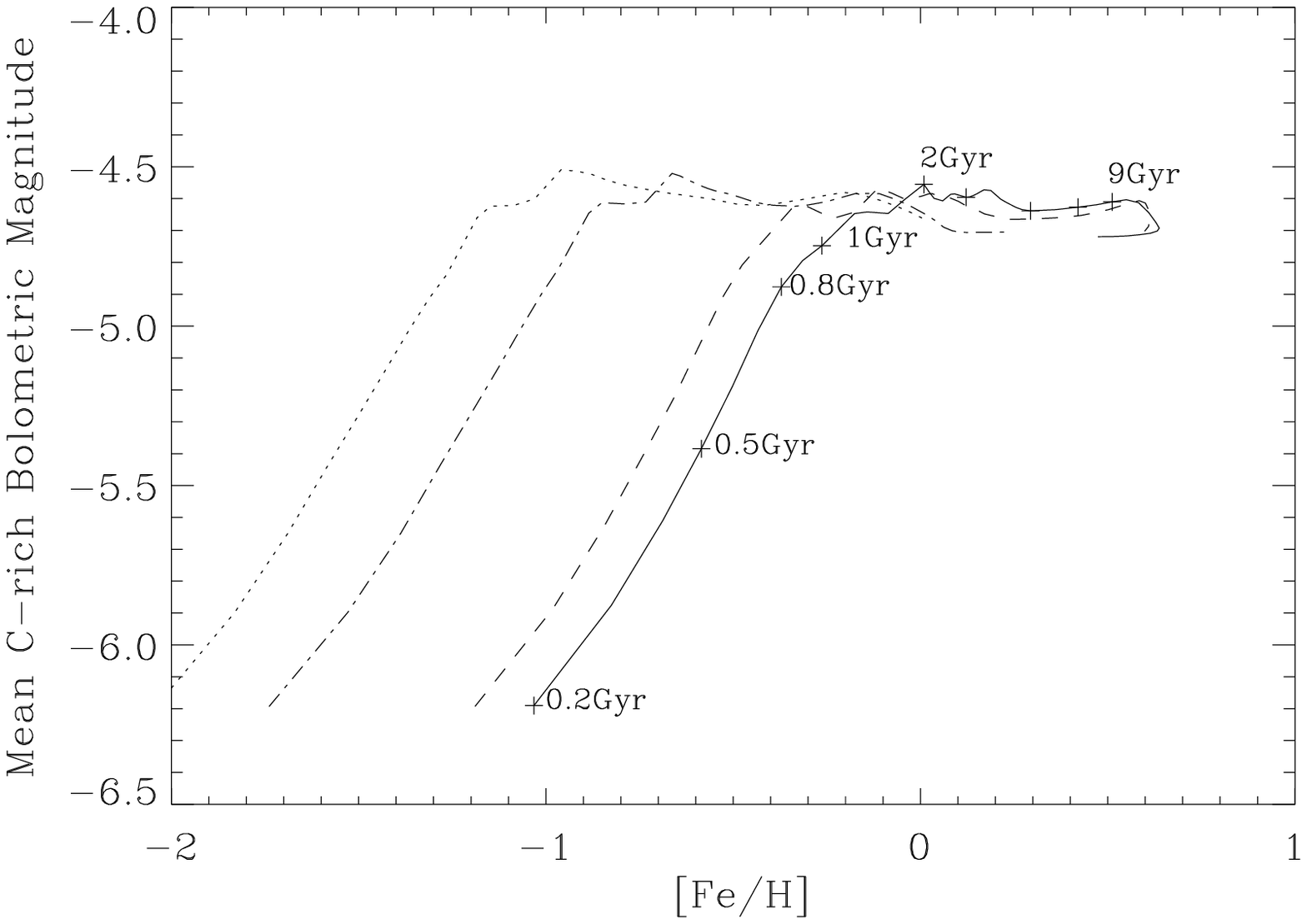}
\includegraphics[clip=,angle=0,width=0.45\textwidth,height=9.cm]
{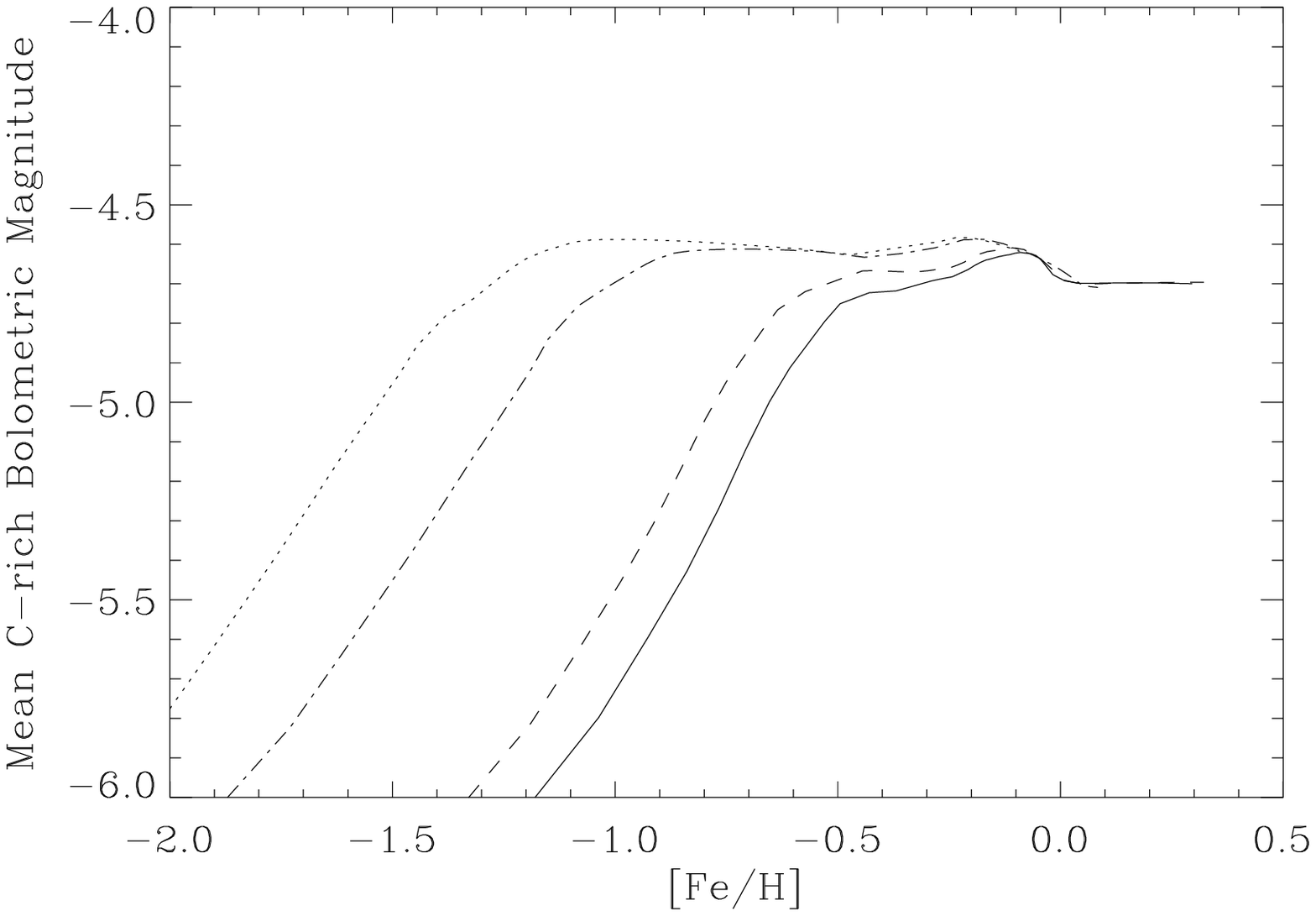}
\caption{Left: Evolution of the mean bolometric magnitude of carbon rich stars 
as function of the parent stellar population metallicity assuming closed box to 
calculate the chemical evolution model. For stellar population older than 
$0.5-0.8$\,Gyr, the mean bolometric magnitude of carbon rich stars is constant 
for wide range of metallicity, up to solar (see text). Right: Similar to the 
left panel but assuming the stellar populations are formed by gas infall }
\label{mean_bol_mag}
\end{figure*}


Figure \ref{mean_bol_mag} shows the evolution of $<\!M_{Bol,C}\!>$ with 
metallicity, for the star formation histories and galaxy formation scenarii 
considered in this paper. A few reference stellar population ages are 
indicated. 

The evolution of $<\!M_{Bol,C}\!>$ as a function of [Fe/H] can be 
summarized as follows. For stellar populations younger than $\sim$1\,Gyr, 
carbon star populations become fainter as time evolves. As already pointed 
out, young systems are dominated by the first generation of chemically 
homogeneous carbon stars. As the turn-off mass of this first
generation decreases, the final stellar core mass and hence the final
luminosity of the predominant carbon stars decreases
(Aaronson \& Mould 1980, Bressan et al. 1994, Mouhcine \& Lan\c{c}on, 2002). 
The shift between the rising curves for various star formation
timescales is due to the differing rates of metal enrichment
of the galaxy gas; the intermediate mass stars forming out of this 
enriched gas have not yet reached the carbon rich phase. 

For stellar systems older than $\sim$0.8-1\,Gyr, when carbon star 
formation has reached the quasi-stationary regime, the evolution of 
$<\!M_{Bol,C}\!>$ is completely 
insensitive to metallicity. As the stellar populations 
grow older and more evolved chemically, (i) the carbon star sequence 
in the HR diagram becomes less extended which reduces the total luminosity
of the carbon star population, (ii) the number of carbon stars decreases. 
As both the total luminosity and the total number 
of carbon stars scale with the metallicity, the ratio is relatively 
independent of it.  

As in the case of N$_{C}$/N$_{M5+}$, the model adopted 
for the galaxy formation (infall or closed box) does not affect 
the evolution of $<\!M_{Bol,C}\!>$ with galaxy metallicity dramatically,
except that the chemical evolution for stellar systems formed via gas infall 
is slower.  The ``plateau" in $<\!M_{Bol,C}\!>$ is thus 
reached at lower metallicity.

The predicted values of $<\!M_{Bol,C}\!>$ for stellar systems with 
stationary carbon star 
formation lie between -4.6 and -4.7. Comparison with the observed  values
in Local Group galaxies with good carbon star statistics and good evidence
that the observed carbon stars are indeed on the TP-AGB (e.g. galaxies that
display an extended giant branch), shows very good agreement: as recalled
in Sect.\,\ref{Lbolc.obs.sec}, observational values lie around -4.7.  

This result leads some investigators to consider carbon star populations as 
potential distance indicators, better in some respects than the Cepheids 
for example (higher bolometric luminosity, single epoch of observation, 
reduced interstellar reddening with near-infrared observations, low 
contamination from dwarf field stars). 
Determinations of the distance modulus to M31 by Brewer et al. (1996) 
and Nowotny et al. (2001) using this technique are in good agreement 
with other distance determinations (Freedman \& Madore1990).

\subsection{The normalized carbon star number as a function of metallicity}

Combining the number count model and the chemically-consistent spectrophotometric 
model, we can predict the relation between metallicity and the number of 
carbon stars normalized to the total stellar luminosity. This relation in 
important in the sense that it enables us to distinguish between the two 
competing explanations for the correlation between N$_{C}$/N$_{M5+}$ and 
metallicity (see Sect.\,\ref{NcNm.obs.sec}).

\begin{figure*}
\includegraphics[clip=,angle=0,width=0.45\textwidth,height=9.cm]{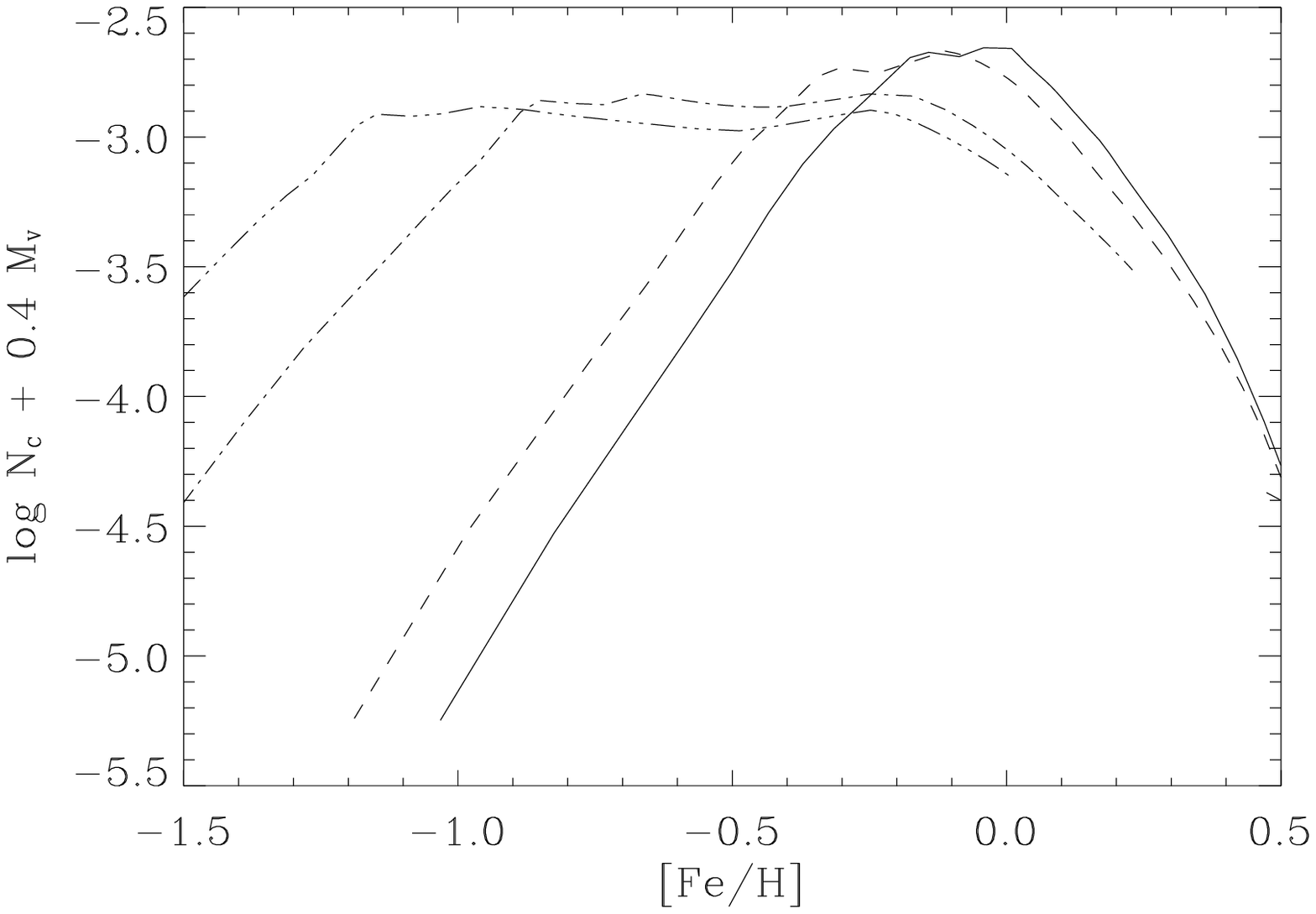}
\includegraphics[clip=,angle=0,width=0.45\textwidth,height=9.cm]{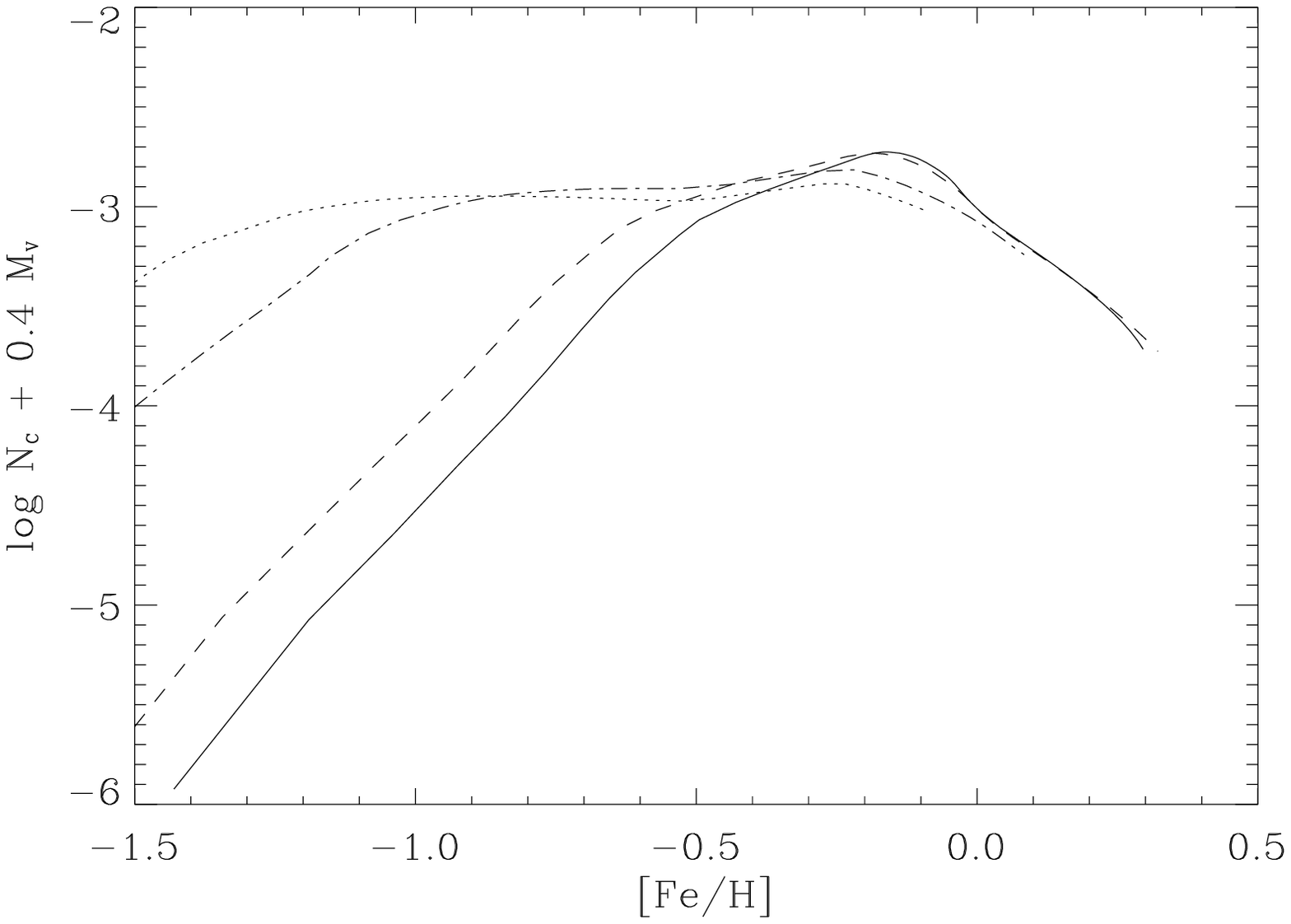}
\caption{Left: Correlation of V-band luminosity-normalized carbon rich star number, 
with the metallicity for closed box model. Right: Similar to the left panel but 
assuming gas infall to form the stellar population.}
\label{NcMv}
\end{figure*}


Figure \ref{NcMv} shows the evolution of the carbon star number normalized 
to the V-band luminosity of the parent galaxy, as a function of metallicity, 
for our standard evolutionary scenarii. 
The evolution of the carbon star number normalized to the B-band light is 
similar, they are obtained from the previous ones by a shift of 0.4\,(B-V), 
which remains small at all times.

Again, the evolution of the normalized number of carbon stars as 
function of metallicity can be split up into two regimes. 
For stellar populations younger than about 1\,Gyr, independently 
of the star formation timescale, carbon stars 
with lower initial masses and correspondingly longer carbon
rich lifetimes accumulate. In the meantime the star formation
timescale sets the rate of increase of the interstellar [Fe/H].

For older stellar systems that have reached the quasi-stationary 
regime of carbon star production, the normalized number of carbon 
stars remains constant over a wide range in metallicity, up to solar. 
This is the case for the same reasons invoked above to interpret 
the insensitivity of the mean bolometric luminosity to the metallicity.
Over this range in metallicity, chemically consistent evolutionary 
models of stellar populations predict linear correlation between 
a galaxy's metallicity and its V (or B) magnitude. 
The plateau in both panels of Fig.\,\ref{NcMv}, together with the 
luminosity-metallicity correlation, illustrate that the number of 
carbon stars is inversely proportional to [Fe/H] over this range 
of ages, even for systems with complex star formation histories. 
A striking feature is that the predicted value of N$_{C,L}$ on 
this plateau is again relatively independent of the star formation 
history. 
This is because the luminosity-metallicity relation is independent 
of the star formation history, as long as latter is continuous.
The predicted average value of the normalized number of carbon star 
is $\sim\,-3$. The agreement between the predicted value and the 
observed ones is rather good.

For old metal-rich systems (i.e. when [Fe/H]$\geq\,$0.), 
all star formation scenarii predict that the normalized number 
of carbon stars declines with increasing metallicity.
This results from the very low efficiency of carbon star 
formation in metal-rich systems and from the fact
that the magnitude-metallicity relation of the galaxies 
is not linear any more. Additional observations 
are needed to obtain more constraints on the carbon star
populations of metal-rich galaxies.

In summary, the details of the star formation 
history do not affect the systematic behavior of the predicted 
relation between the metallicity 
and the normalized number of carbon stars, as long as 
a quasi-stationary regime has been reached. 

\section{N$_{C}$/N$_{M}$ as an abundance indicator}
\label{NcNm_Zindic.sec}

The above investigations of the relations between the properties 
of carbon star populations and the parent galaxy metallicity 
have lead us to highlight the relatively short time that separates 
the birth of the first stars from the appearance of the bulk of 
the carbon stars. This time is indeed short when compared
to the timescales over which the average SFR varies in Hubble 
sequence galaxies. Therefore, for stellar systems older than 
$\sim$\,1\,Gyr, for which carbon star production proceeds at 
a quasi-stationary rate, the N$_{C}$/N$_{M}$ ratio indicates 
metallicity rather than age. 

The time delay between the production of the bulk of the carbon stars, 
in terms of number, and the ejection of the bulk of iron to the interstellar 
medium (which is related to the lifetime of SNe Ia progenitors since
SNe Ia events are the main producers of iron) is also short in comparison to 
the star formation timescale. The difference between the initial metallicity 
of progenitors of carbon stars and the interstellar medium metallicity 
when those stars become carbon rich is not significant, except
at very early times in the life of a galaxy.  
This means that N$_{C}$/N$_{M}$ ratio measures both the metallicity 
of the carbon star progenitors and the present interstellar medium 
metallicity.

This suggests that the N$_{C}$/N$_{M}$ radial profile, for stellar systems 
that form stars continuously, will follow the metallicity profile and both 
will have the same shape (i.e. slope). This statement is valid at least for 
stellar systems for which the evolution is not affected by processes that 
could trigger rapid radial mixing of the interstellar material 
(such as bars or merging events for example), with a timescale shorter 
or comparable to the timescale of the chemical evolution, and hence carbon 
star production. Bars or interactions could significantly
alter the local composition of the interstellar medium
on timescales shorter than 1\,Gyr, and break the relation between
this composition and N$_{C}$/N$_{M}$.

Observations supporting this results already exist in the literature. 
Indeed Brewer et al. (1995) have imaged five fields along the semi-major 
axis of M31 to derive the radial profile of N$_{C}$/N$_{M}$. Using the 
compilation of Pritchet et al. (1987) to calibrate the N$_{C}$/N$_{M}$ 
vs. [Fe/H] relation, and the observed N$_{C}$/N$_{M}$ radial profile, 
they have derived the radial abundance profile along the semi-major axis 
of M31. They find rather reasonable agreement 
with the metallicity gradient observed by Blair et al. (1981, 1982) 
and Dennefeld \& Kunth (1981) using HII regions and supernova remnants.

Using N$_{C}$/N$_{M}$ as an alternative abundance indicator has 
interesting advantages. It enables us to investigate the 
chemical evolution in environments where the traditional methods 
are useless, for instance far out in disks where HII regions are rare,
or in the halo of galaxies. The knowledge of abundance gradient at large 
galactic scales can be of great help to constrain 
models of galaxy formation.

\section{Summary and conclusions}
\label{Concl.sec}

In this paper we have provided a theoretical investigation
of carbon star populations and of related statistical properties 
as a function of metallicity, and we have confronted these models 
with available observations in Local Group galaxies. To achieve 
this goal we have constructed evolutionary synthesis models that 
use a large grid of stellar evolution tracks, including the TP-AGB.
The underlying TP-AGB models take into account the processes that 
determine the evolution of those stars in the HR diagram and 
establish the duration of the TP-AGB phase. To provide predictions 
for the nature of the cool and luminous intermediate age stellar 
content of stellar populations (carbon rich or oxygen rich stars, 
optically visible or dust-enshrouded ones), 
our models account explicitly for the metallicity dependence 
of the evolution of TP-AGB stars. We thus derive estimates
of the carbon star formation efficiency and its evolution
as a function of metallicity for large grid of metallicity 
from Z/Z$_{\odot}$\,=\,1/50 to Z/Z$_{\odot}$\,=\,2.5.b
Note that because fundamental physical processes that govern 
TP-AGB evolution remain poorly understood and are calibrated 
in the solar neighbourood or the Megellanic Clouds, uncertainties 
grow when moving away from this metallicity domain. 

The evolution of carbon star properties as a function of metallicity 
for stellar systems with continuous star formation was modeled via 
new chemically consistent population synthesis models. Those models
use metallicity-dependent stellar yields available in the literature 
for massive stars, and our synthetic TP-AGB evolution models for
the new metallicity-dependent yields of intermediate mass stars. 
The effect of SNe Ia on the chemical evolution is also included. 
Effect of gas infall and of the star formation history were explored.

Comparisons between predicted carbon star population statistics as 
a function of the metallicity of the interstellar medium
and the available data show that our models are, for the first time, 
able to reproduce qualitatively and quantitatively the observations
in the Local Group galaxies. This success supports the choices made
in the inputs of the evolutionary models. The models show that the 
evolution of the properties of carbon star populations are established 
by a combination of the following effects:
\begin{itemize}
\item The temperature of the giant branch varies with metallicity, 
leading to more late type M stars with increasing metallicity.  
\item In metal-poor systems, dredge-up events are more efficient
in producing carbon stars and carbon rich lifetimes are longer.
This is due to (i) higher core masses at the onset of the TP-AGB 
phase and the resulting earlier onset of third dredge-up events, 
and (ii) longer interpulse periods leading to more violent third 
dredge-up events.   
\item The time needed for a stellar population to form the 
bulk of its carbon stars ($\sim 1$\,Gyr) is significantly shorter 
than the typical evolutionary timescale along the Hubble sequence. 
This means that for systems older than $\sim 1$\,Gyr
the evolution of the statistical properties of carbon star populations 
will merely reflect the sensitivity of the evolution of individual
TP-AGB stars to metallicity. 
\end{itemize}

The last effect implies that the observed statistics are mainly 
determined by the current metallicity.
Star formation history or any other process which has a timescale 
longer than about 1\,Gyr have no significant effects on the evolution 
of the statistics of carbon stars as function of metallicity. 
This means that the observed N$_{C}$/N$_{M5+}$ vs. [Fe/H] correlation 
is a metallicity sequence rather than an age sequence. 
Consequently, we predict that, at least for unbarred spirals or 
noninteracting galaxies, the radial profile of N$_{C}$/N$_{M5+}$ ratio 
will have the same slope as the radial abundance profile. 

We have also given estimates on the fraction of carbon stars that 
may be missed in optical surveys, showing that at the age when the 
bulk of carbon stars are formed ($\sim$0.8 Gyr after the burst), 
only 10\%-20\% are dust-enshrouded. We argue that the statistics 
of carbon star populations will not be affected dramatically by 
the missed high mass-loss rate stars. 

The models show that, for stellar systems older than $\sim\,0.8-1\,$Gyr, 
carbon star populations have the following properties: 

\begin{itemize}
\item The mean bolometric luminosity of the carbon stars is independent 
of metallicity. Such a behavior over a wide range of metallicity 
is consistent with a long record of observational constraints discussed 
in the literature (Aaronson \& Mould 1985, Richer et al. 1985). 
The value derived from our calculations is $<\!M_{Bol,C}\!>\,=\,-4.7$. 
Carbon stars can be considered as potential distance indicators.
\item The number of carbon stars normalized to the luminosity of the 
parent stellar system is independent of metallicity over a wide range 
in abundance. This behavior supports 
the interpretation of the anti-correlation between N$_{C}$/N$_{M5+}$ 
and [Fe/H] as due, partially, to more efficient carbon star formation 
at lower metallicity. The value derived from our calculations is 
$\log(N_{C,L})\,\simeq\,-3$. 
This is consistent with the observational constraints discussed 
recently by Azzopardi et al. (1999), where the authors report that the 
observed value is independent of metallicity and dispersed around -3.3.
\end{itemize}
We note that our models were able to reproduce the observed statistics 
considering only the carbon stars formed via the third dredge-up channel, 
which means that faint, dwarf carbon stars will represent a small 
number fraction of the whole population of carbon stars.   

\section{Acknowledgments}
We are particularly grateful to J. K\"{o}ppen, for helpful interactions 
and enlightening discussions reg. chemical evolution.

\bsp

\label{lastpage}

\end{document}